\documentclass[a4paper,11pt]{article}
\pdfoutput=1 % if your are submitting a pdflatex (i.e. if you have
\usepackage{jheppub} % for details on the use of the package, please
\usepackage[T1]{fontenc} % if needed
\usepackage{enumerate}
\usepackage{subfigure}
\usepackage{tabularx}

\newcommand{\mtop}{m_{\mathrm{top}}}

\title{\boldmath Effects of color reconnection on $t\bar{t}$ final states at the LHC}

\author[a,b]{Spyros Argyropoulos}
\author[a]{and Torbj\"orn Sj\"ostrand}

\affiliation[a]{Department of Astronomy and Theoretical Physics, Lund University, S\"olvegatan 14A, SE-223 62 Lund, Sweden}
\affiliation[b]{Deutsches Elektronen Synchrotron, Notkestra\ss e 85, D-22607, Hamburg, Germany}

\emailAdd{spyridon.argyropoulos@desy.de}
\emailAdd{torbjorn@thep.lu.se}

\abstract{The modeling of color reconnection has become one of the dominant sources of systematic uncertainty in the top mass determination at hadron colliders. The uncertainty on the top mass due to color reconnection is conventionally estimated by taking the difference in the predictions of a model with and a model without color reconnection. We show that this procedure underestimates the uncertainty when applied to the existing models in {\sc Pythia}~8. We introduce two new classes of color reconnection models, each containing several variants, which encompass a variety of scenarios that could be realized in nature and we study how they affect the reconstruction of the top mass. After tuning the new models to existing LHC data, the remaining spread of predictions is used to derive a more realistic uncertainty for the top mass, which is found to be around 500 MeV. We also propose how future LHC measurements with $t\bar{t}$ events can be used to further constrain these models and reduce the associated modeling uncertainty.}

\begin{document} 
\begin{flushright}
LU TP 14-23\\
DESY 14-134\\
MCnet-14-15\\
July 2014
\end{flushright}
\maketitle
\flushbottom

\section{Introduction}\label{sec:intro}
The measurements of the top quark mass at hadron colliders are becoming more and more precise, with the most recent ones achieving a precision of less than 1 GeV ~\cite{topMass_Combination}. Interactions and interference between the top decay products during hadronization, also known as color reconnection (CR), decrease the precision that can be achieved and constitute 20 to 40\% of the quoted top mass uncertainty. The modeling of CR is therefore a non-negligible limiting factor to the precision that could be achieved \cite{topMass_overview}.

Color reconnection is an ad hoc mechanism aiming to describe the interactions that can occur between chromoelectric fields during the hadronization transition. This is expected to occur at a significant rate at the LHC due to the high number of colored partons, from a combination of multiparton interactions (MPI) and parton showers as well as from the beam remnants. Top quarks produced at the interaction point will travel a distance $\ell=\gamma\beta c\hbar/\Gamma_{\mathrm{top}}$ in the detector before they decay. The typical transverse mass of top quarks at the LHC corresponds to a boost factor of $\gamma\simeq1.5$ and a transverse decay length of $\ell\simeq0.2$ fm, which is smaller than the typical hadronization scale of 1 fm. It is thus possible that the top quarks will interact with the color force fields that are stretched between the colored partons in the final state. 

Due to the lack of observables and measurements that could constrain these effects in $t\bar{t}$ final states, the CR uncertainty is estimated by taking the difference $\Delta\mtop=\mtop^{\mathrm{CR}}-\mtop^{\mathrm{no-CR}}$ between the reconstructed top mass in a model with CR and in a model without CR, but with the same input top mass. Since there are no first-principles models that could give a unique answer, the best that one can do is to study a range of not unrealistic models and evaluate the spread of effects.

Up until now, most of these studies have been carried out with \textsc{Pythia}~6 \cite{Pythia6}, which featured several CR models. The \textsc{Pythia}~8 \cite{Pythia8} successor has not had a corresponding range of models, but only one. Extending the above procedure for the calculation of $\Delta m_{\mathrm{top}}$ to \textsc{Pythia}~8 would raise some concerns. Notably, the default CR setup in \textsc{Pythia}~8 does not directly affect the top decay products, so it is likely that it underestimates the effect of CR on the top mass. Furthermore, since it is known that models without CR cannot describe minimum-bias data \cite{Perugia,CMS-CR}, $\mtop^{\mathrm{no-CR}}$ corresponds to a scenario which is not realized in nature. One should consider instead the spread between realistic CR models in order to obtain a more reliable estimate for the top mass uncertainty.

These issues point to the need for a systematic study of how CR specifically affects the top decay products in a hadronic collision. The goal of the study presented here is threefold. First we introduce two new classes of CR models in \textsc{Pythia} 8, extreme ones that are designed to affect only the top decay products but in a maximal way, and full-scale ones where CR can affect all colored partons in the final state. With this broader collection of models we study the effect of CR on the top mass and explain the different ways in which these models affect the reconstruction of the top system. Finally, we propose observables that could be measured at the LHC in order to eliminate or constrain these new models, providing a better control of the associated uncertainty.

This paper is organized as follows. The new CR models are described in section \ref{sec:models}. Section \ref{sec:reco} describes the generation and reconstruction of the $t\bar{t}$ sample that we use for this study. Section \ref{sec:top} discusses the effects of the new CR models on the top mass and analyzes the way in which each model affects the reconstruction of the $t\bar{t}$ final state. The tuning of these models to LHC data is discussed in the same section. Lastly, section \ref{sec:observables} presents a series of observables that are sensitive to CR and could be used to constrain or even eliminate these new CR models.

\section{Description of the models}\label{sec:models}

Since the introduction\footnote{For a brief historical account of the development of the related ideas see \cite{Sjo2013}.} of CR \cite{vanZijl} as a mechanism to explain the increase of the average transverse momentum as a function of the charged particle multiplicity, observed by UA1 \cite{UA1}, several CR models have been developed and incorporated in \textsc{Pythia}~6. In addition to the simple baseline \cite{CR-SS}, these include variants of the color annealing \cite{CRannealing,CR-SW}, generalized area law \cite{CR-GAL} and soft color interaction (SCI) \cite{CR-SCI} models. In the second and third set of models, CR is viewed as a modification of the geometry of the string configuration that minimizes a certain measure such as the invariant length of a string piece (color annealing) or the invariant mass of a pair of partons (area law), while in the SCI model, CR is described as a stochastic color exchange between the perturbative partons and a background color field. Although sharing common traits, the models implemented in {\sc Pythia}~8 are not identical with the ones in \textsc{Pythia}~6,  and therefore extend the range of available models. Models for CR are also available in other general-purpose event generators like {\sc Herwig++} \cite{CR-Herwig} and {\sc Sherpa} \cite{CR-Sherpa}.

Generally, CR can be viewed as a combination of static and of dynamical effects, with different emphasis among models. The static effects are typically associated with `accidental singlets', whose occurrence is dictated by the $SU(3)$ multiplet structure. For instance, considering the top decay products, there is a $1/9$ probability that the $b$ quark has the same color as the $W$ decay products. Therefore, if the space-time separation between the $W$ and the $b$ is small, it is physically possible that the string spanned between the $q\bar{q}$ pair reattaches to the $b$ quark `by accident'. From a more dynamical point of view, the color strings that are stretched out when partons fly apart are made up out of (infinitely) many colored soft gluons. An overlap of two such strings thus involves many potential soft exchanges and interactions, in all possible color states. Therefore the geometrical overlap rather than the color algebra may drive the color reconnection probabilities \cite{khozeww}. 

The old and new models implemented in \textsc{Pythia}~8, to be described here, do not provide a fully dynamical framework nor a complete treatment of the color multiplet structure of the final state. In these models, the occurrence of a reconnection is determined by other criteria, which do assume underlying dynamical mechanisms and therefore allow reconnection probabilities to go up to unity. 

\textsc{Pythia}~8.1 currently contains only one model (sometimes referred to as the MPI-based one), with two possibilities for resonant systems, hereafter denoted as `default' and `default Early Resonance Decays (ERD)'. Color reconnections in the default model is a two-step procedure proceeding as follows:
\begin{enumerate}
\item Starting from the lowest-$p_T$ interaction in a set of multiple parton interactions, a reconnection probability for an interaction with hardness scale $p_T$ is calculated by
\begin{equation}\label{eq:recProb}
P_{\mathrm{rec}}(p_T)=\frac{(R_{\mathrm{rec}}\ p_{T0})^2}{(R_{\mathrm{rec}}\ p_{T0})^2+p_T^2},
\end{equation}
where $0\le R_{\mathrm{rec}}$ ($\le 10$, since by then effects have saturated) is a phenomenological parameter and $p_{T0}$ is an energy dependent parameter used to damp the low-$p_T$ divergence of the $2\rightarrow 2$ QCD cross section. The reconnection probability is chosen to be higher for soft systems, reflecting that the latter are described by more extended wave functions, thus having a higher probability to overlap and interact with other systems. If an interaction does not reconnect with the next-highest one in $p_T$, then consecutively higher ones are tried, so that the total reconnection probability for an interaction is $1 - (1 - P_{\mathrm{rec}})^n$ if there are $n$ interactions at higher $p_T$ scales.
\item The interactions where a reconnection will happen are sorted in decreasing $p_T$. Starting from the interaction with the highest $p_T$, all color dipoles $(i,j)$ are found. Color is described in the $N_C \to \infty$ limit, so that each quark corresponds to the end of one color dipole, whereas a gluon corresponds to two dipole ends, one for its color and another for its anticolor. The softer interactions which should be reconnected are studied iteratively. For each such interaction, the gluons $k$ are inserted (in decreasing $p_T$) into the dipole that minimizes the increase in the $\lambda$ measure\footnote{The notation $\lambda_{k;ij}$ signifies the increase in the length of the string spanned between partons $i$ and $j$ due to the addition of a gluon kink $k$.} \cite{lambda} 
\begin{equation}\label{eq:lambdaoldmodel}
\Delta\lambda = \lambda_{k;ij} \equiv \lambda_{ik}+\lambda_{jk}-\lambda_{ij}=\ln\frac{(p_i\cdot p_k)(p_j\cdot p_k)}{(p_i\cdot p_j) m_0^2}.
\end{equation} 
Also $q\bar{q}$ pairs originating from the splitting of a gluon are handled in a similar spirit, whereas (the few) other quarks are not affected.
\end{enumerate}
The $\lambda$ measure is a generalization of the rapidity range available for particle production. For the case of a $q\bar{q}$ string with endpoints $i$ and $j$, the rapidity range is given by $\Delta y=\ln\left(m_{ij}^2/m_0^2\right)$, where $m_0$ is a typical hadronic mass scale, say of the order of the $\rho$ mass. A string with $n$ gluon excitations between the quark and antiquark (or diquark) is comprised of $n+1$ pieces and the total rapidity range becomes
\begin{equation}\label{eq:lambda1}
\lambda=\sum_{\substack{i=1\\j=i+1}}^{n+1}\ln\frac{m_{ij}^2}{m_0^2}=\ln\prod_{\substack{i=1\\j=i+1}}^{n+1}\frac{m_{ij}^2}{m_0^2}.
\end{equation}
The total $\lambda$ of an event is calculated by summing (\ref{eq:lambda1}) over all of the strings in it.

In the Lund model, $\lambda$ is proportional to the string length. Since $\lambda\sim\Delta y\sim \langle n\rangle$, minimizing the string length $\lambda$ is equivalent to minimizing the average hadron multiplicity $\langle n\rangle$. Loosely speaking, $\lambda$ can be viewed as the `free energy' of a string system, available for particle production. It is generally assumed that, other things being the same, Nature would prefer a low string length \cite{vanZijl}. It should be noted that such a principle does not apply to the perturbative stage of an event, where the hard interaction and MPIs signal the transition from a state of small $\lambda$ (partons confined in the incoming protons) to a state of significantly higher $\lambda$. The principle of string length minimization refers rather to longer timescales, when strings begin to be pulled out between the partons moving out from the central collision. 

The $\lambda$ expression in eq.~(\ref{eq:lambda1}) is actually a simplification of the correct measure, valid when all invariant masses are large. In the limit that a gluon becomes soft it would seem that the attached string pieces could give an arbitrarily much negative contribution to the total $\lambda$. The complete formula \cite{lambda} avoids this problem, but is rather cumbersome to work with. Instead, for the models to be introduced later, a simple modification
\begin{equation}\label{eq:lambda2}
\lambda_{ij} = \ln\frac{m_{ij}^2}{m_0^2} \to \ln \left( 1 + \frac{m_{ij}^2}{m_0^2} \right)
\end{equation}
is introduced to avoid an unphysical behavior when $m_{ij}^2 \to 0$. This modification is not required in eq.~(\ref{eq:lambdaoldmodel}), since an $E_k \to 0$ (for fixed relative angles) would affect all $\lambda_{k;ij}$ the same way and thus not alter the choice of winning $ij$ dipole. 

So far we have not addressed top events specifically. When zooming in on tops, it is relevant to consider the time scales involved. At LHC energies the incoming protons are strongly Lorentz contracted, and thus all the hard interactions take place within a time shorter than 0.001~fm. Related initial- and final-state radiation stretches out to time scales inverse to the hardness of the emission considered, times a time dilatation factor. Viewed in the rest frame of a top, this leads to negligible interference between perturbative activity inside and outside the top system at energy scales above $\Gamma_t$
\cite{khoze}. This is at around the lower shower cutoff scale, and thus color interference effects can be neglected at the perturbative level, in particular relative to the nonperturbative effects that we study here. 

When the top decays after $\sim 0.2$~fm, thus all multiparton interactions and most shower activity in the rest of the event has already occurred, and strings are forming between the receding partons. The subsequent $W$ decays are yet somewhat later. This is still smaller than the 1~fm size scale of hadrons and strings, which is also approximately the typical invariant time for the string fields to disappear by the formation of hadrons. Color reconnections can thus occur between the late-starter partons from the top decays and the partons from the rest of the event, but the space-time difference is still sufficiently large that reconnections may be modeled differently between the two. 

In the existing CR model, the default `late resonance decays' option implements the extreme point of view that the top lifetime is enough to shield the top decay products from color reconnection with the rest of the event. Only the top quark itself is involved in the CR machinery, and the top decays are carried out after the CR have taken place. There is also no CR among the top decay products. In particular, this means that the $b$ quarks are always connected to the same color partners as their top ancestors, and the $W$ bosons will decay to color connected $q\bar{q}$ dipoles that will not participate in the CR. 

This restriction is lifted in the `default ERD' model. Here the tops and $W$s are allowed to decay before the CR machinery is invoked, so the top decay products directly participate in CR. Thus gluons from softer interactions can be attached to the dipoles spanned by the $b$ quarks and by the $q\bar{q}$ decay products of the $W$. 

These two alternatives may be viewed as extemes in the case of the top, but can be fully appropriate for other particles; the Standard Model Higgs is so long-lived that it is immune to reconnection effects, whereas squarks and gluinos can be very short-lived.

The default model is constructed with the guiding principle of perturbing the perturbatively defined `hard' dipoles as little as possible, whereas softer gluons are more easily rearranged. This implies in particular that the ends of the top-decay dipoles are fixed by the perturbative color flow, and CR can only contribute by inserting gluon kinks on the hadronizing strings. 

In order to probe the full extent of the effects that CR can have on $t\bar{t}$ final states, we introduce a new class of models that is designed to affect the top decay products separately from the rest of the reconnections. In these models, first the default framework (with `late resonance decays') is used to carry out the normal CR, at this stage involving the undecayed top. Then the $t$ and $W$ are allowed to decay with the decay products eventually radiating, and finally the resulting new partons are allowed to reconnect with the rest of the event. This second top-reconnection step will be done using a variety of choices, as described below. Given that the starting point is the default CR model, this class of models does not require any re-tuning in order to describe minimum bias data, which simplifies studies. The choice of allowing separate CR rules for the underlying event and for the top decay products can be justified in two ways. Firstly, there is the time difference, whereby the top decay products are dumped into an underlying-event environment already partly formed. Secondly, the default CR model has been rather little tested and constrained by studies of hard physics at scales close to the top mass, leaving some leeway in the modeling there.  

\begin{figure}[h]
\centering
\centerline{\includegraphics[height=3cm]{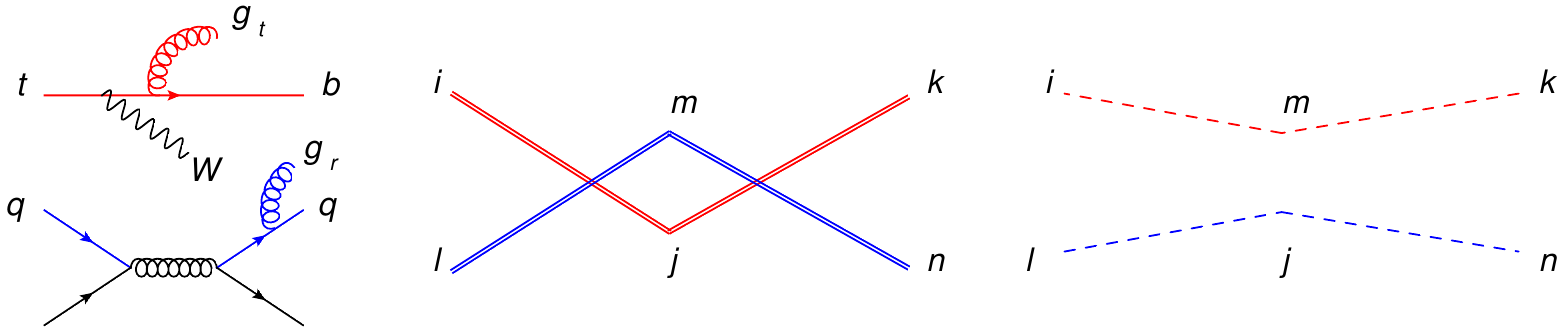}}
\caption{An example of a gluon emitted from a top decay product $g_t$ exchanging colors with a gluon from another interaction $g_r$. A possible string configuration for this final state is shown on the right. The gluons $g_t,g_r$ are identified with the kinks $j,m$ respectively, while the top and bottom quarks and the incoming and outgoing $q$ from the additional interaction are identified with the string endpoints $i,k,l$ and $n$ respectively. The double lines indicate the configuration that results from the perturbative color flow and the dashed lines indicate a configuration with a smaller $\lambda$, resulting from the exchange of colors and anticolors between gluons $j$ and $m$.} 
\label{fig:CR}
\end{figure}

Technically, the reconnections involving the top decay products are done as follows. Two collections of gluons are constructed, one containing the gluons radiated from the top decay products $\{g_t\}$ (excluding radiation from the top quark itself before it decays\footnote{Gluon radiation from the top quark is assigned to $\{g_r\}$.}) and the other containing the gluons from the rest of the event $\{g_r\}$. Iterating over $\{g_t\}$ in random order, one forces the gluons from the top to exchange colors with a gluon from the rest of the event. This color exchange is done in various ways, leading to the following models:
\\\\
\begin{tabularx}{\linewidth}{@{\quad\quad}l@{\qquad}X@{}}
  1. forced random & $g_t$ is forced to exchange colors with a random gluon from the $\{g_r\}$ set,  \\
  2. forced nearest & $g_t$ is forced to exchange colors with the $g_r$ that minimizes $m(g_t,g_r)^2=(p(g_t)+p(g_r))^2$, \\
  3. forced farthest & $g_t$ is forced to exchange colors with the $g_r$ that maximizes $m(g_t,g_r)^2$, \\
  4. forced smallest $\Delta\lambda$ & a gluon $j\in\{g_t\}$ color-connected to partons $i$ and $k$ is forced to exchange colors with the gluon $m\in\{g_r\}$ (color-connected to partons $l$ and $n$) for which the $\lambda$ change
  \begin{equation}\label{eq:dLambda}
  \Delta \lambda(j,m) =\lambda_{m;ik}+\lambda_{j;ln}-\left(\lambda_{j;ik}+\lambda_{m;ln}\right)
 %= \lambda_{im} + \lambda_{mk} + \lambda_{lj} + \lambda_{jn} -
  %\left( \lambda_{ij} + \lambda_{jk} + \lambda_{lm} + \lambda_{mn}  \right) ,
  \end{equation}
 is minimal,\\
 5. smallest $\Delta\lambda$ & as in the previous model, except that gluons exchange colors only if $\Delta\lambda<0$.
\end{tabularx}
\\\\
An illustration of the effect of a color exchange between gluons is shown in figure~\ref{fig:CR}. As we shall see, some of these options can give unrealistically large effects, so a phenomenological strength parameter $\alpha$, $0 \leq \alpha \leq 1$, can be used to reduce effects: each parton in the $\{g_t\}$ set is only tested for reconnection with probability $\alpha$.

We note that allowing also quarks and antiquarks to reconnect in the above models leads to small quantitative changes in some distributions that are sensitive to CR effects, while leaving the qualitative picture described in these studies unchanged. In order to reduce complexity, we therefore restrict CR only among gluons in this class of models. Another feature of these models is that gluons from the rest of the event are allowed to participate in reconnections multiple times. Notably this implies that a low (high) momentum gluon in the forced nearest (farthest) scenario can lead to color swapping between the top decay products, through multiple color exchanges. Variants could be considered where a gluon would be allowed to exchange colors only once, or where the distance measure is in angle rather than mass. In view of such considerations, the forced random model may be the most relevant of the first three, given that an outgoing parton from the top decay encounters partons and fields from the rest of the event in a more-or-less random manner. Notably, MPIs are spread across the transverse collision area, and produce partons moving in random directions, so spatial location does not equal momentum direction at early times.

A second class of models is also introduced, which performs reconnections between all gluons in the event, irrespective of the type of process where they are produced. This class contains two models dubbed `swap' and `move'. The swap model is a generalization of the `smallest $\Delta\lambda$' model described above. For each final-state gluon pair $j$ and $m$ one calculates the difference in $\lambda$ resulting from the exchange of the color and anticolor indices, using eq.~(\ref{eq:dLambda}). A reconnection is performed if $\min_{j,m}\Delta\lambda(j,m)\leq\Delta\lambda_{\mathrm{cut}}$, where $\Delta\lambda_{\mathrm{cut}}\leq 0$ is a tunable parameter that expresses a CR strength. The procedure is repeated until no allowed swaps remain. As in the previous set of models, it is possible to restrict the CR effects by considering only a fraction $\alpha$ of the gluons.

\begin{figure}[h]
\centering
\centerline{\includegraphics[height=3cm]{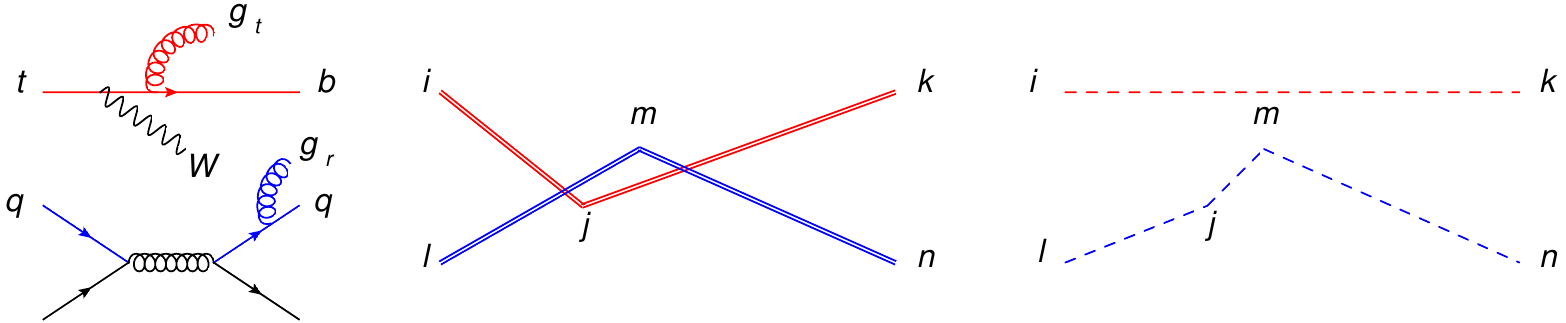}}
\caption{In the `move' model, a gluon $j$ originally attached to string piece $ik$ can be moved to a different string piece $lm$ if it leads to a smaller total string length $\lambda$. Solid lines indicate the original configuration and dashed lines indicate the resulting configuration after moving the gluon.} 
\label{fig:CR2}
\end{figure}

The `move' model works as follows. Starting from a final state gluon $j$ attached to a string piece between partons $i$ and $k$, the change in the string length $\Delta\lambda$ that would result from moving the gluon to another string piece between partons $l$ and $m$ is calculated using
\begin{equation}
\Delta\lambda(j, lm) = \lambda_{j;lm}-\lambda_{j;ik}=
\lambda_{lj} + \lambda_{jm} + \lambda_{ik} - 
\left( \lambda_{ij} + \lambda_{jk} + \lambda_{lm} \right).
\end{equation}
Having considered all final state dipoles, gluon $j$ is finally moved to the dipole for which $\min_{j,lm}\Delta\lambda(j,lm)\leq\Delta\lambda_{\mathrm{cut}}$. The procedure is iterated for all final state gluons, or a fraction $\alpha$ thereof. A graphical illustration is displayed in figure \ref{fig:CR2}. Moving a gluon away from a two-gluon singlet is forbidden, since that would give a leftover singlet gluon. 

\begin{figure}[h]
\centering
\centerline{\includegraphics[height=3cm]{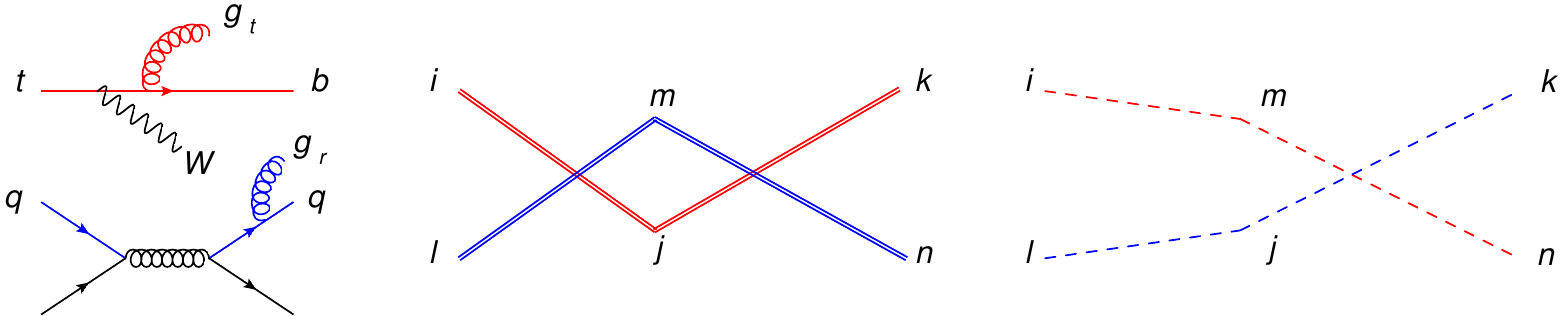}}
\caption{Illustration of the effect of the `flip' model. The same process as in figure \ref{fig:CR}, shown with the underlying string configuration. The solid lines indicate the initial configuration and the dashed lines represent a flip in the string pieces $ij$ and $lm$, resulting from the exchange of one of the color indices between gluons $j$ and $m$. The figure represents a case where a flip reduces the total string length $\lambda$. We note that after the flip, the $b$ quark from the top decay (endpoint $k$) is color connected to quark $l$ from a separate MPI.} 
\label{fig:CRflip}
\end{figure}

Neither `swap' nor `move' reconnect quarks. That is, if a $q\bar{q}$ pair start out at the opposite ends of a string then so they will remain. In the `swap' model, the gluons found along this string can change, and in the `move' model even the number of such gluons, but the endpoints not. To lift this restriction, such that e.g.\ the $b$ can hook up with the $\bar{q}$ from the $W$ decay, a `flip' step can be added subsequent to the `swap' or `move' procedure. The basic idea here is to flip two string pieces, from $i$ to $j$ and from $l$ to $m$, so that colors instead go from from $i$ to $m$ and from $l$ to $j$ (figure \ref{fig:CRflip}). For any two color singlet system one finds the set $(ijlm)$ for which a flip would reduce the total $\lambda$ the most. Then $\min_{ij,lm}\Delta\lambda(ij,lm)=\min_{ij,lm}\left[\lambda_{im}+\lambda_{lj}-(\lambda_{ij}+\lambda_{lm})\right]\leq\Delta\lambda_{\mathrm{cut}}$ is selected for a flip. Here singlet systems that have undergone one flip are not allowed any further ones, or else the procedure leads to the formation of many low-mass $gg$ systems, thus markedly reducing the charged particle multiplicity. 
While the normal string has a color and an anticolor at opposite ends, there is also the possibility of junction topologies, where three quarks are at the ends of a Y-shaped field configuration. As a minor variation, such topologies are either excluded or included among the allowed flip possibilities. 

Since the `swap' and `move' models affect all scattering processes, they have to be tuned using minimum bias data. This is described in the following section.

\section{Generation and reconstruction of $t\bar{t}$ final states}\label{sec:reco}
The studies presented in this paper have been performed with simulated $t\bar{t}$ events, generated with \textsc{Pythia} version 8.185 at a center-of-mass energy of $\sqrt{s}=8$ TeV. \textsc{Pythia} provides leading order matrix elements for $q\bar{q}\rightarrow t\bar{t}$ and $gg\rightarrow t\bar{t}$. On top of the $t\bar{t}$ process, \textsc{Pythia} attaches initial and final state parton showers and multiparton interactions, which evolve from the scale of the hard process down to the hadronization scale in an interleaved manner \cite{Cor10a}. The generation of $t\bar{t}$ events was done using the leading order PDF set CTEQ6L1 \cite{cteq6l1} with the 4C tune \cite{Cor10a}. Particles with a proper decay length of $c\tau>10$ mm were considered stable. The default parameters of \textsc{Pythia} 8.185 were used for the particle properties, in particular the top mass and width were set to $\mtop=171$ GeV and $\Gamma_{\mathrm{top}}=1.4$ GeV.

The top quark almost always decays via $t\rightarrow Wb$, with the $W$ further decaying into a lepton and a neutrino or a quark-antiquark pair. Final states containing a $t\bar{t}$ pair are thus classified in three categories according to the $W$ decays: all hadronic, where both $W$'s decay hadronically, semi-leptonic, where one $W$ decays hadronically, and dilepton, where both $W$'s decay into leptons. Dilepton events are expected to be the least sensitive to CR effects and suffer from the presence of two neutrinos in the final state, which complicate the reconstruction of the top. Increasing the number of hadronic $W$ decays corresponds to an increased probability for CR but also to an increased combinatorial complexity in the reconstruction of the top. In this study we restrict ourselves to semi-leptonic $t\bar{t}$ events, which offer a balance between these two effects. Semi-leptonic final states are also used in the currently most precise measurements of the top mass \cite{topMass_Combination}. We further consider only $W$ decays to electrons and muons, in order to reduce the difficulties associated with the reconstruction of $\tau$ leptons.

The selection and reconstruction of $t\bar{t}$ events is inspired by what is done experimentally, however it doesn't aim to reproduce the level of sophistication of a full-fledged experimental analysis. The event is reconstructed starting from stable electrons and muons with a transverse momentum $p_T>25$ GeV and pseudorapidity $|\eta|<2.5$. The missing transverse energy (MET) is reconstructed by summing the four-momenta of all neutrinos within $|\eta|<5$. Jets are reconstructed from final-state stable particles, lying within $|\eta|<4.5$, using the anti-kt algorithm \cite{anti-kt} with a radius parameter of 0.4. Neutrinos are excluded from the jet clustering. The jets are separated into $b$-jets and light flavor jets by requesting the presence of at least one $B$ hadron within the radius of the jet or the absence thereof. These $B$ hadrons must lie within $|\eta|<2.5$ and have $p_T>5$ GeV. Only jets with $p_T>25$ GeV and $|\eta|<2.5$ are considered in the reconstruction of the top. Jets that are within $\Delta R<0.4$ from electrons or muons are not considered in the analysis.

The event selection is performed separately in the electron and muon channel starting from the inputs described above. In the electron channel we request the presence of exactly 1 electron, at least 2 light flavor jets and at least 2 $b$-jets. The hadronically decaying $W$ is reconstructed from the pair of light flavor jets whose invariant mass is closer to the expected $W$ mass $m_W=80.4$ GeV. Events with $\mathrm{MET}<35$ GeV or $m_T(W)=\sqrt{m_W^2+p_T^2(W)}<25$ GeV are rejected. The selection in the muon channel is the same, except for the cuts on MET and $m_T(W)$, which are replaced by the combined requirement that $\mathrm{MET}+m_T(W)>60$ GeV. 

For the reconstruction of the top system, the two $b$-jets with the highest $p_T$ are combined with the hadronically and leptonically decaying $W$ candidates and the combination which minimizes the difference $m_{t}-m_{\bar{t}}$ is selected. The leptonically decaying $W$ boson candidate is formed by the lepton and the transverse component of the sum of the four-momenta of the neutrinos. The analysis is implemented in the Rivet framework \cite{rivet}.

\section{Effects of color reconnection on $t\bar{t}$ observables}\label{sec:top}
Here we present the effects that the different CR models have on observables related to the $t\bar{t}$ final state. 

\subsection{Top mass}
The top mass is reconstructed in semi-leptonic events as outlined in section \ref{sec:reco}, using only the hadronically decaying top candidate, whose constituents' four-momenta can be experimentally fully reconstructed. Here we consider only those events where the mass of the hadronically decaying $W$ candidate is in the range $75\leq m_{W}\leq85$ GeV and where all the jets that constitute the hadronic top candidate have $p_T>40$ GeV. To obtain an estimator for the top mass in each ensemble of pseudo-data we perform a one-dimensional fit to the $\mtop$ distribution using a Gauss distribution, defined as
\begin{equation}\label{fit}
G(m;\sigma)=C\cdot\frac{\mathrm{e}^{-(m-\widehat{m}_{\mathrm{top}})^{2}}}{2\sigma^2},
\end{equation}
where $C$ is a floating normalization. The estimator for the top mass $\widehat{m}_{\mathrm{top}}$ is obtained by a $\chi^2$ minimization fit using Minuit \cite{minuit}. We note that in the quest for ultimate precision, current experimental analyses usually employ more sophisticated methods for the extraction of the top mass, which may have a different sensitivity to CR than the method described here. Nevertheless, the simple one-dimensional fit that we use here should adequately capture the underlying physics effects and provide a reliable order-of-magnitude estimate for the CR uncertainty on the top mass.

\begin{figure}[h]
\centering
\centerline{\includegraphics[width=0.49\columnwidth]{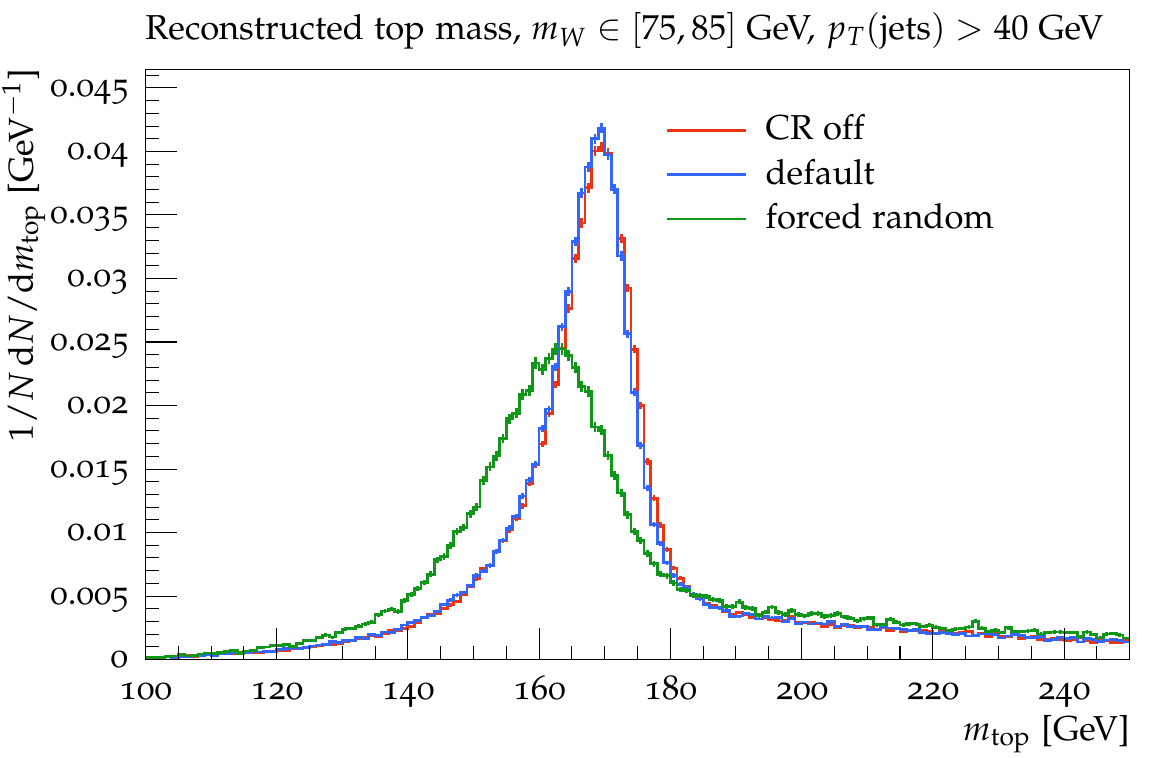}
\includegraphics[width=0.49\columnwidth]{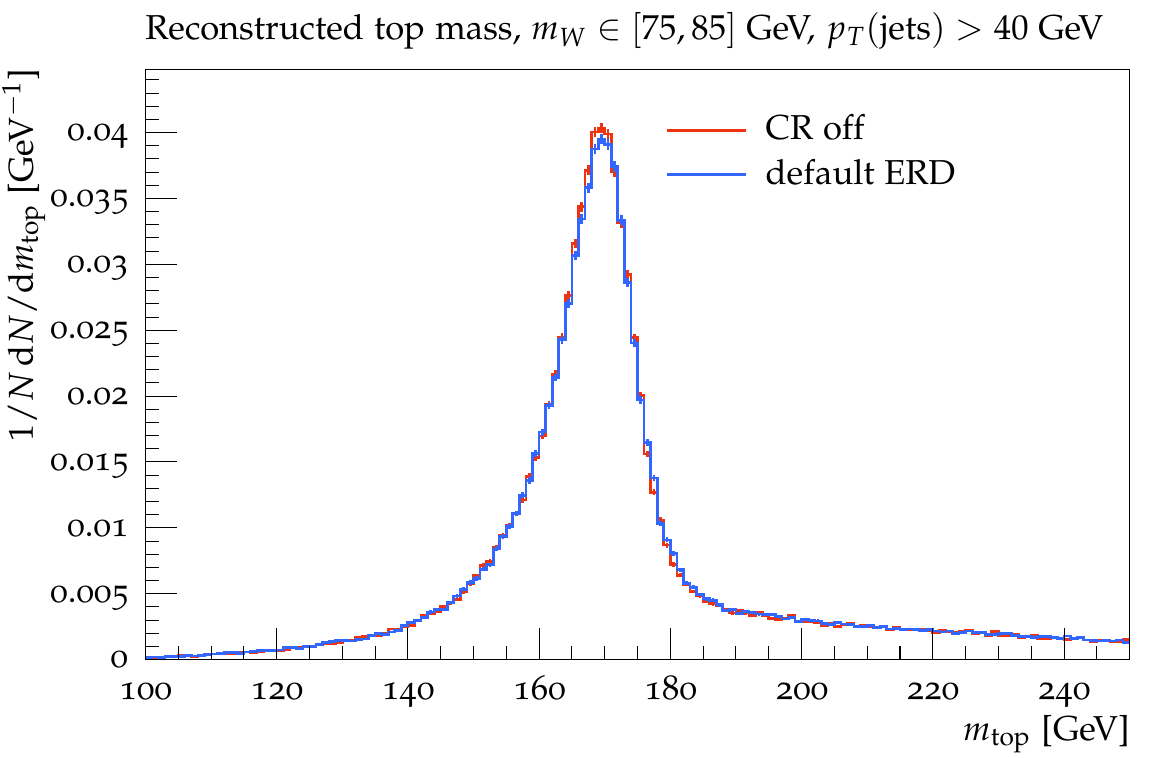}}
\caption{Most of the models predict a shift of $\mtop$ to lower values (left) but there are models that predict a positive shift (right). Models that force the top decay products to reconnect have large effects (left).} 
\label{fig:mtop}
\end{figure}

Since CR affects all jets in the same way, one can use the well-known $W$ mass to correct for systematic shifts in the energy scale of the jets \cite{CR-SW}. For each pseudo-data ensemble, we define a Jet Energy Scale (JES) correction factor $R_{\mathrm{JES}}=80.385\mathrm{\ GeV}/\widehat{m}_{W}$, where $\widehat{m}_{W}$ is obtained by fitting the invariant mass distribution of the two jets that form the hadronic $W$ candidate. A rescaled top mass estimator is then defined by $\widehat{m}_{\mathrm{top}}^{\mathrm{rescaled}}=R_{\mathrm{JES}}\cdot\widehat{m}_{\mathrm{top}}$. The results for the different CR models are collected in Table \ref{table:fits} and figure~\ref{fig:mtop} shows the $\mtop$ distributions for some of the standard and outlier models. These were obtained starting from tune 4C, but choosing a maximal reconnection strength ($R_{\mathrm{rec}}=10$), in order to probe how big an effect CR can have on the top mass.
\begin{table}[tbp]
\centering
\begin{tabular}{|lccc|}
\hline
Model & $\widehat{m}_{\mathrm{top}}$ [GeV] & $\Delta \widehat{m}_{\mathrm{top}}$ [GeV]  &$\Delta \widehat{m}_{\mathrm{top}}^{\mathrm{rescaled}}$ [GeV]\\
\hline\hline 
CR off						& $169.377\pm 0.034$ & 0                                  & 0\\
default						& $168.962\pm 0.024$ & $-0.415\pm 0.042$ &  $+0.209\pm0.065$  \\
default ERD					& $169.758\pm 0.063$ & $+0.381\pm 0.072$ &  $+0.285\pm0.092$  \\
forced random					& $162.407\pm 0.062$ & $-6.970\pm 0.071$  &   $-6.508\pm0.142$ \\
forced nearest					& $166.507\pm 0.037$ & $-2.870\pm 0.050$  &   $-1.439\pm0.093$ \\
forced farthest					& $163.630\pm 0.050$ & $-5.747\pm 0.060$  &   $-3.193\pm0.109$ \\
forced smallest $\Delta\lambda$  	& $167.646\pm 0.027$ & $-1.731\pm 0.043$  &   $-0.754\pm0.081$ \\
smallest $\Delta\lambda$			& $168.649\pm 0.031$ & $-0.728\pm 0.046$  &   $+0.156\pm0.074$  \\
swap						& $169.169\pm 0.030$ & $-0.208\pm 0.045$  &    $-0.051\pm0.087$ \\
move						& $169.349\pm 0.037$ & $-0.028\pm 0.050$  &   $-0.028\pm0.086$ \\
swap+flip (excl. junc.)			& $169.119\pm 0.035$ & $-0.258\pm 0.049$  &   $+0.065\pm0.086$ \\
swap+flip (incl. junc.)			& $169.040\pm 0.034$ & $-0.337\pm 0.048$  &   $-0.047\pm0.082$ \\
move+flip (excl. junc.)			& $169.244\pm 0.037$ & $-0.133\pm 0.050$  &   $-0.036\pm0.088$ \\
move+flip (incl. junc.)			& $169.160\pm 0.035$ & $-0.217\pm 0.049$  &   $+0.040\pm0.089$ \\
\hline\hline
\end{tabular}
\caption{\label{table:fits}Values of $\mtop$ predicted by the different CR models. The difference $\Delta \widehat{m}_{\mathrm{top}}=\widehat{m}_{\mathrm{top}}-\widehat{m}^{\mathrm{no-CR}}_{\mathrm{top}}$ before rescaling is given in the middle column and after rescaling in the rightmost column.}\label{table1}
\end{table}
\\\\\\
We observe that
\begin{enumerate}[i.]
	\item one can construct models that have a big impact on the top mass reconstruction; in particular models that enforce color reconnection tend to give a big shift in $m_{\mathrm{top}}$, while models based on a minimization principle lead to smaller shifts, as expected,	
	\item color reconnection can produce both positive and negative shifts in $\mtop$, with positive shifts being smaller in magnitude than negative shifts,
	\item the difference $\widehat{m}_{\mathrm{top}}^{\mathrm{default}}-\widehat{m}_{\mathrm{top}}^{\mathrm{no-CR}}$ does not cover the range of top mass values predicted by the different CR models. 
\end{enumerate}
The question that arises is what produces the shifts in $\mtop$ and why the different CR models have so different behavior. We recall that the top mass is reconstructed from the four-momenta of a $b$-jet and the light flavor jets that come from the $W$ decay, i.e. $\mtop^2=(p(b)+p(j_1)+p(j_2))^2$. Therefore shifts in $\mtop$ can originate from two distinct mechanisms: i) changes in the four-momenta of the individual jets, e.g. due to leakage of hadrons out of the jet cone, and ii) changes in the kinematic correlations between the three constituent jets, e.g. the $b$-jet can become more collinear to the light jets. 

\begin{figure}[h]
\centering
\centerline{\includegraphics[width=0.49\columnwidth]{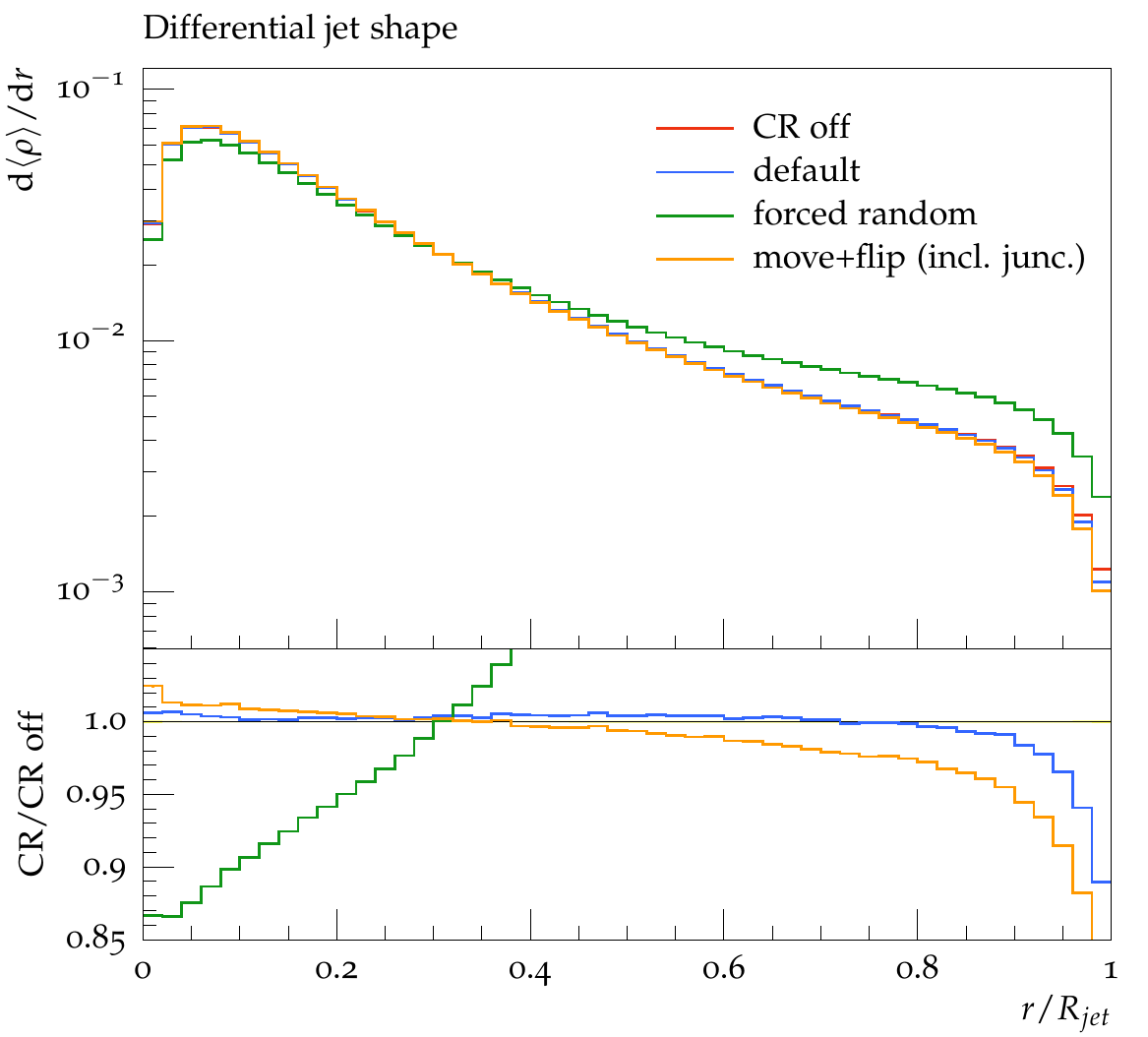}
\includegraphics[width=0.49\columnwidth]{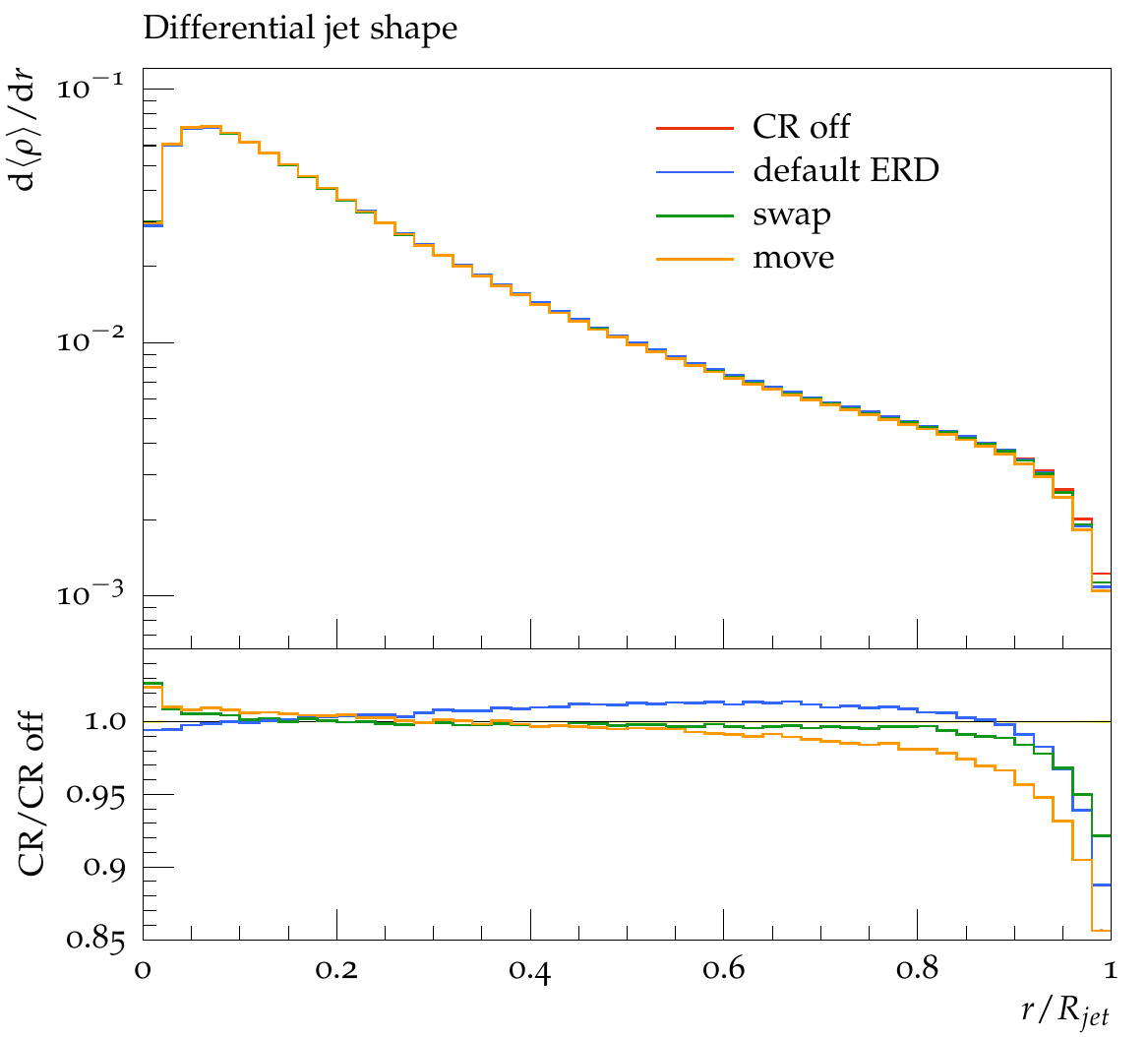}}
\caption{Models of CR that lead to a negative (positive) mass shift usually predict broader (narrower) jets (left). For some of the models this is not the case (right).} 
\label{fig:shape}
\end{figure}

We study the first mechanism by reconstructing the differential jet shape
\begin{equation}\label{eq:diffshape}
\rho(r)\equiv \sum_{i\in\mathrm{jet}}\frac{E_i\left(r-\frac{\Delta r}{2},r+\frac{\Delta r}{2}\right)}{E_{\mathrm{jet}}},
\end{equation}
which gives the fraction of the jet's energy contained in an annulus of inner radius $r-\Delta r/2$ and outer radius $r+\Delta r/2$, as a function of the distance $r=\sqrt{(\Delta\eta)^2+(\Delta\phi)^2}$ from the jet axis. The differential jet shape $\rho(r)$ averaged over all jets in the simulated events is plotted as a function of the distance from the jet axis in figure \ref{fig:shape}. We observe that, in most cases, models that predict a negative (positive) top mass shift involve broader (narrower) jets. Therefore, the shift in $\mtop$ can be explained by the loss (or gain) of the hadronic decay products out of the jet cone as a result of the rearrangement of the hadronizing strings. At one extreme, the `forced' models allow the partons of the top jets to be reconnected to many different partons located outside the jet cone, leading to broader jets and a loss of energy. At the other extreme, models that reduce $\lambda$ (e.g. the `default', `swap' and `move' shown in figure \ref{fig:shape}) can give more narrow jets. Since the default color flow in parton showers tends to pick a small $\lambda$ inside jets to begin with, it is more difficult to make jets more narrow than more broad, thereby explaining the asymmetric mass-shift range. The relation between $\lambda$ shift and top mass shift does not hold for all of the models, however, so this mechanism cannot be solely responsible for the top mass shift.

\begin{figure}[h]
\centering
\centerline{\includegraphics[width=0.49\columnwidth]{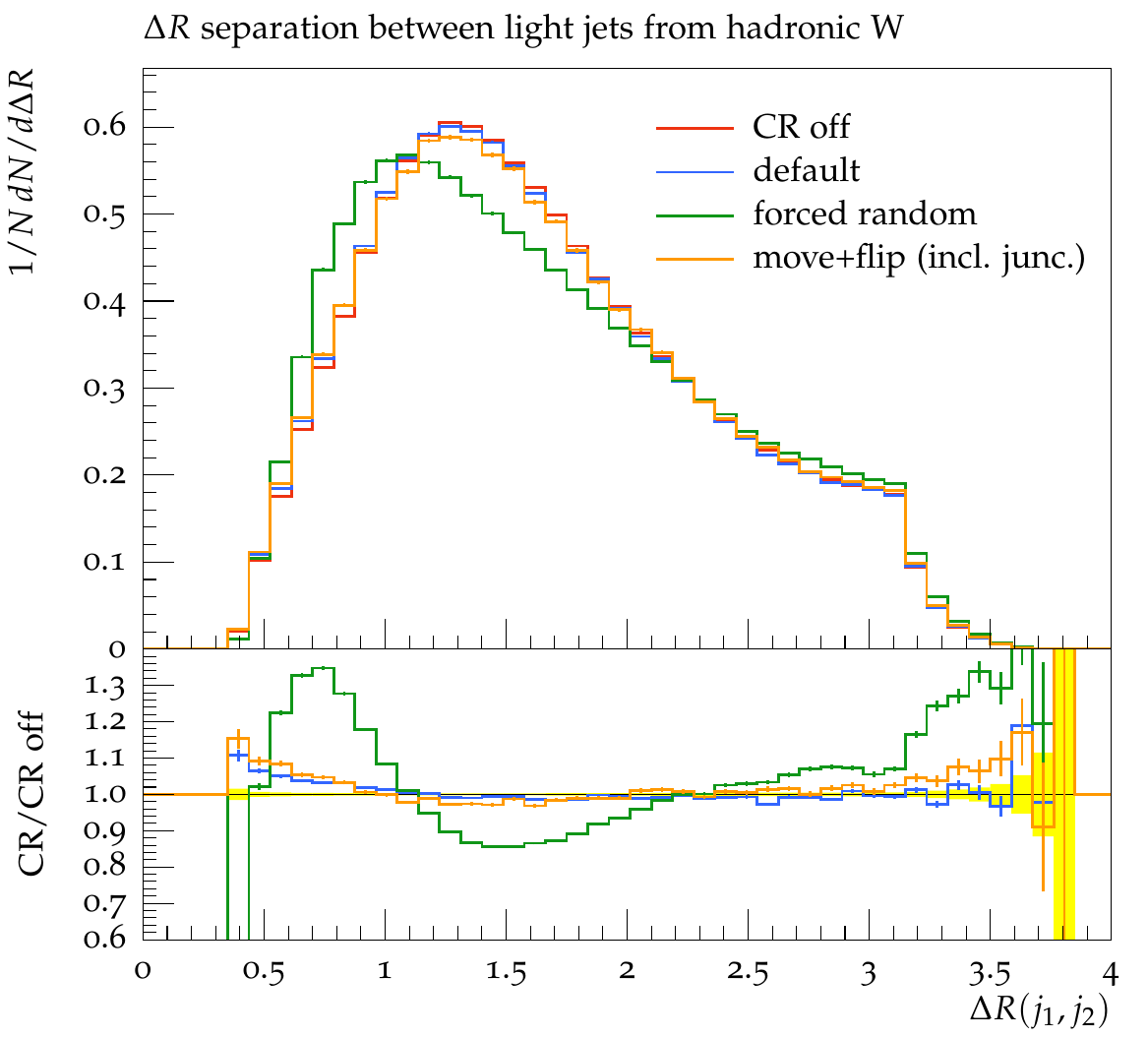}
\includegraphics[width=0.49\columnwidth]{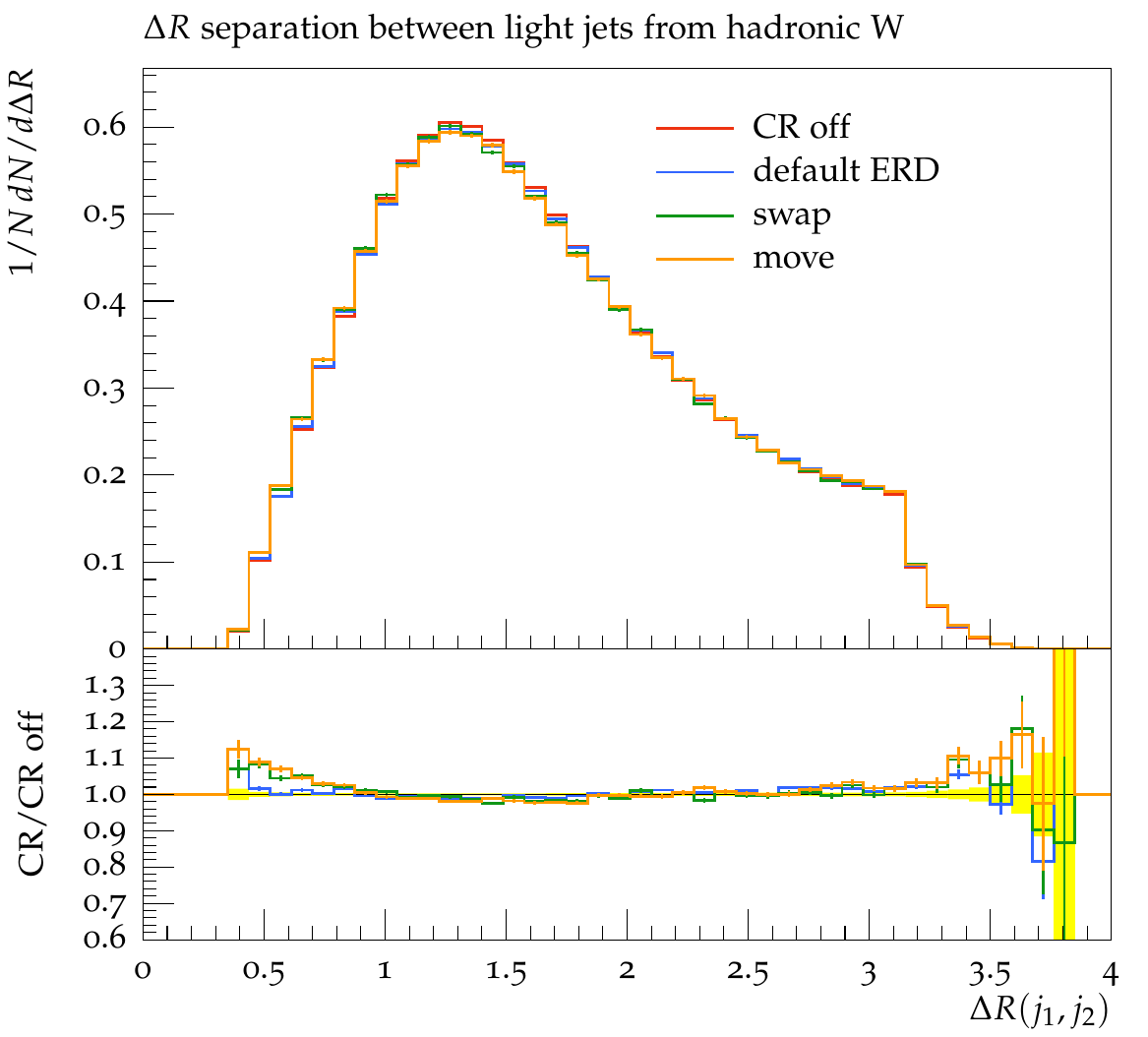}}
\caption{Models of CR that lead to a negative mass shift usually involve more collinear decay products. The direction of the top mass shift is determined by the combined effect of CR on the jet shapes and on the separation of the top decay products  (cf. figure \ref{fig:shape}).} 
\label{fig:deltaR}
\end{figure}

The second mechanism which could affect $\mtop$ is kinematic correlations between the jets formed by the top decay products. These can be probed by the $\Delta R$ separation between the light jets from the hadronically decaying $W$ and between the $b$-jet and the $W$. As shown in figure \ref{fig:deltaR}, the models that predict a smaller top mass generally predict that the $W$ decay products are more collinear. 
%A notable exception is the `move' model, which also leads to more collinear $W$ decay products but results in a positive $\mtop$ shift. It can be seen however (figure \ref{fig:shape}), that this model predicts narrower jets than the model without CR, however. The latter implies that out-of-cone radiation will in general be smaller than the model without CR, resulting in a higher energy for the reconstructed jets, thus driving $\mtop$ to higher values. 
Concerning the separation between the $b$-jet from the top decay and the $W$, we find that in the models which correspond to the highest $\Delta\mtop$, the $b$-jet tends to be closer to the $W$ boson. In the rest of the models, no significant changes in $\Delta R(b,W)$ are observed. A shift of reconstructed jet directions depending on the color flow is a well-known `string effect' \cite{earlystring}, verified experimentally \cite{jadestring,alephstring}.

We conclude that the effect of CR on the top mass is a combination (and in some cases a competition) between two distinct effects: changes in the jet structure and changes in the kinematic correlations between the top decay products. Both effects originate from the rearrangement of the hadronic fragmentation products, as a result of the rearrangement of the hadronizing strings. 

\subsection{Constraints from existing data}

In the previous section we showed that there are CR models which can lead to a big shift in $m_{\mathrm{top}}$. In this section we address the question whether such models can be accommodated by existing data. As we already discussed, jet shapes are sensitive to color reconnection and can thus be used to constrain the class of models that affect $t\bar{t}$ events. Similarly the `swap' and `move' models which affect all final states can be constrained by minimum bias data. We therefore tune both classes of models to $t\bar{t}$ and minimum bias data respectively and revisit their effect on the top mass.

Jet shapes in $t\bar{t}$ have been measured by the ATLAS collaboration at $\sqrt{s}=7$ TeV \cite{ATLASjetshapes}. Models of CR that affect $t\bar{t}$ events are tuned to this data by varying the strength parameter $0\leq \alpha\leq 1$ that determines the fraction of final state gluons radiated from top decay products that take part in the color reconnections. With the rest of the parameters determined by tune 4C \cite{Cor10a}, we find that the best description of the data corresponds to a choice of $\alpha$ between 5 and 10\%. In the following we fix this parameter to $\alpha=7.5\%$.

The `swap' and `move' models that affect all final states, as well as the model without CR are tuned to minimum bias data measured by ATLAS at $\sqrt{s}=7$ TeV \cite{ATLASminbias}. Starting from the tune 4C, the parameters that are varied are $\Delta\lambda_{\mathrm{cut}}$ and $p_{T0}^{\mathrm{ref}}$. The latter is a parameter that appears in the regularization of the cross-section for massless $2\rightarrow 2$ parton scattering as follows. The divergent $p_{T}^{-4}$ behavior of the cross-section is replaced by
\begin{equation}
\frac{1}{p_T^4}\longrightarrow\frac{1}{p_T^4}\frac{p_T^4}{\left(p_{T0}^2(\sqrt{s})+p_T^2\right)^2},
\end{equation}
with
\begin{equation}\label{eq:pt0ref}
p_{T0}(\sqrt{s})=p_{T0}^{\mathrm{ref}}\left(\frac{\sqrt{s}}{\sqrt{s}_{\mathrm{ref}}}\right)^{\sqrt{s}_{\mathrm{pow}}}.
\end{equation}
The parameter $p_{T0}$ in (\ref{eq:pt0ref}) is the same as the one appearing in eq. (\ref{eq:recProb}). In (\ref{eq:pt0ref}), $\sqrt{s}_{\mathrm{ref}}$ is an arbitrary reference energy and $\sqrt{s}_{\mathrm{pow}}$ is a tunable parameter fixed in this study by the 4C tune. We find that for the model without CR, and for the `swap' model, the data are best described by the choice $p_{T0}^{\mathrm{ref}}=2.3$ GeV, while for the `move' model the best description is obtained with $p_{T0}^{\mathrm{ref}}=2.25$ GeV. For the `swap+flip' and `move+flip' scenarios we use $p_{T0}^{\mathrm{ref}}=2.2$ GeV an $p_{T0}^{\mathrm{ref}}=2.15$ GeV respectively.\footnote{For comparison, we remind that in the 4C tune, $p_{T0}^{\mathrm{ref}}=2.085$.} Both the `swap' and the `move' models, require a maximal color reconnection strength to describe data, therefore we set $\Delta\lambda_{\mathrm{cut}}=0$. We note that what we present here is not a comprehensive minimum bias tune, which would involve several parameters, but an effort to get the best possible description of data with the minimal set of modifications to the baseline tune.

\begin{figure}[h]
\centering
\centerline{\includegraphics[width=0.49\columnwidth]{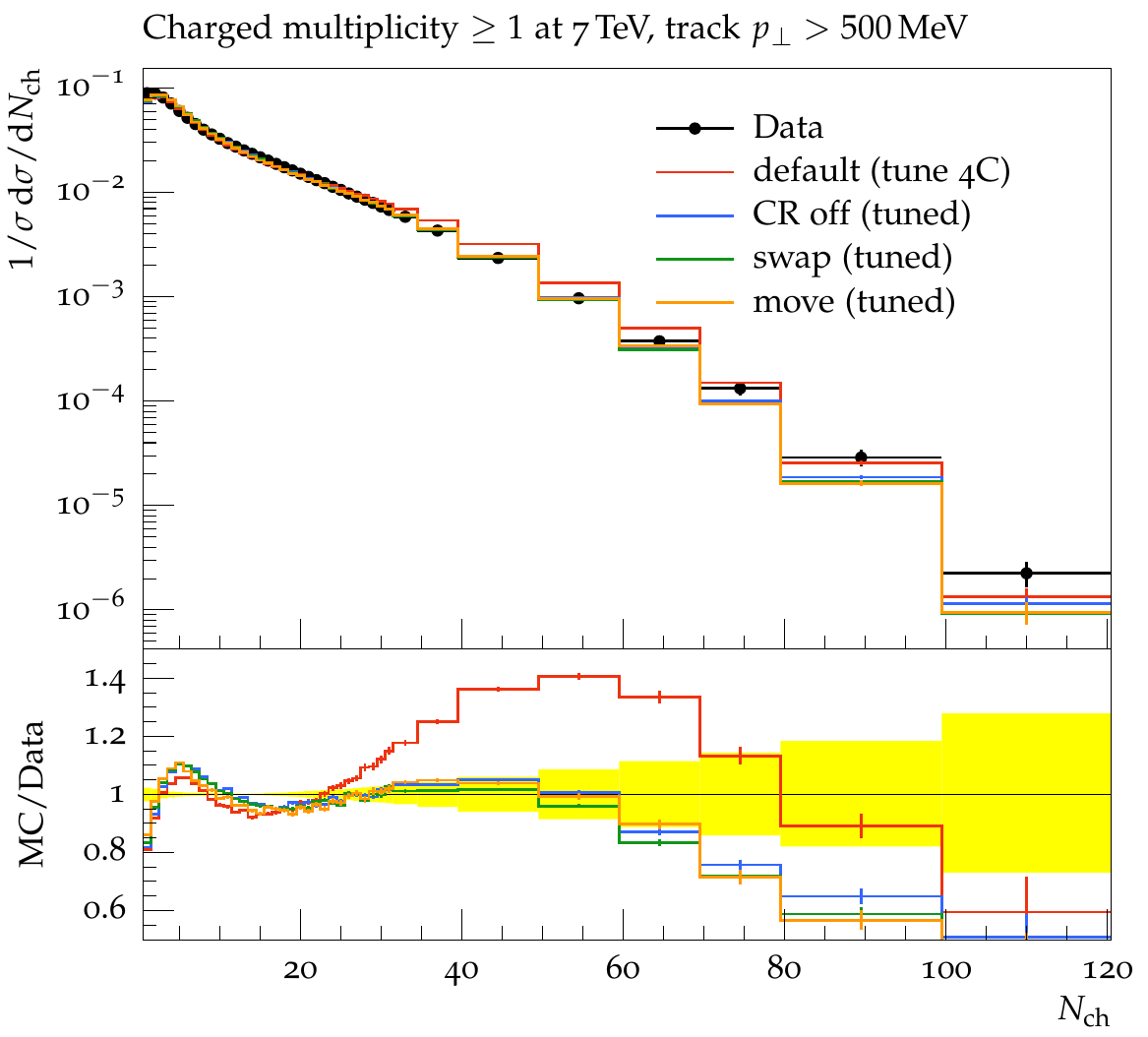}
\includegraphics[width=0.49\columnwidth]{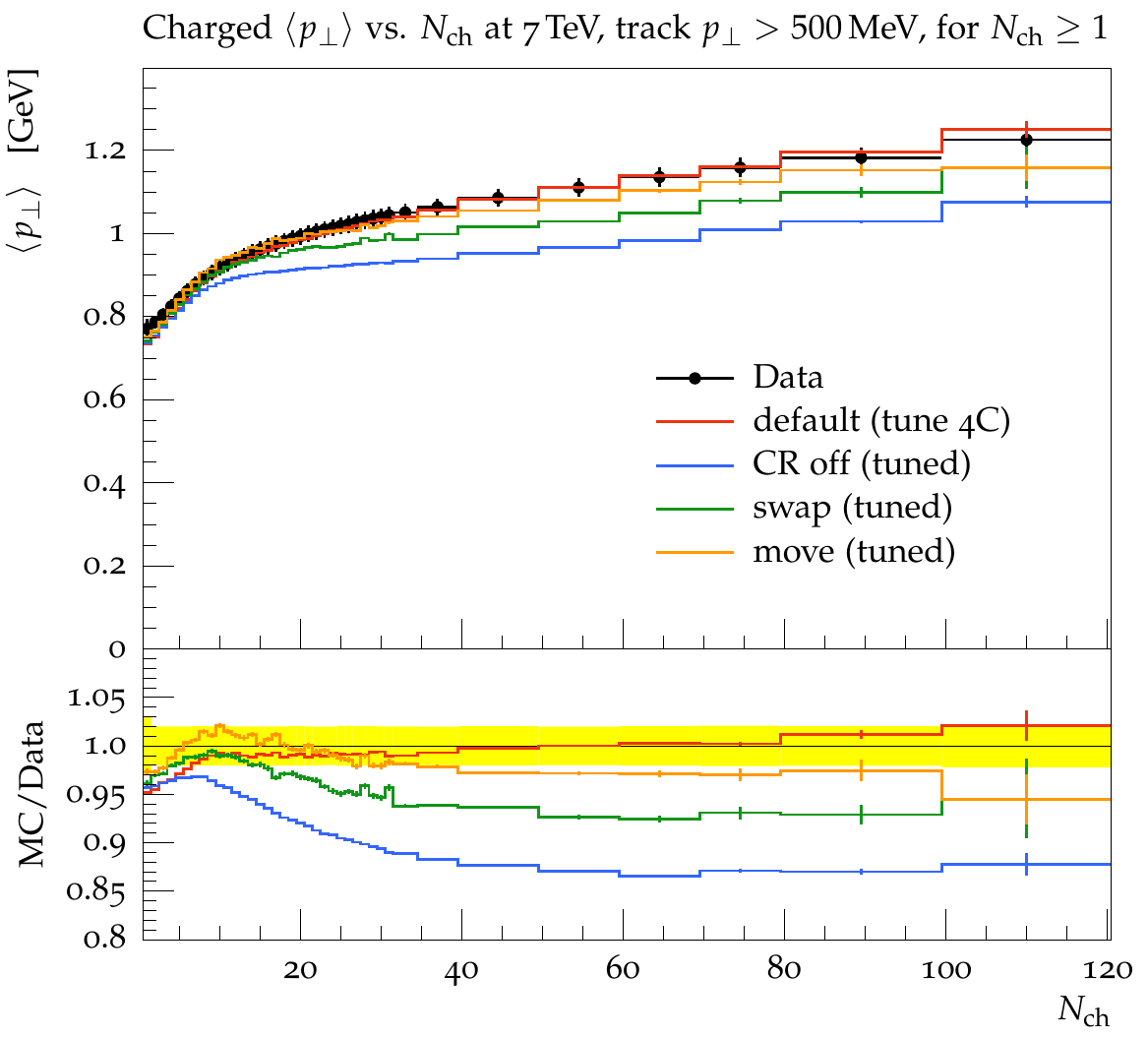}}
\caption{Multiplicity (left) and mean $p_T$ per charged particle (right) for particles with $p_T>500$ MeV in minimum bias events. The plots display the best fit for the default model obtained with tune 4C and for the `CR off', `swap' and `move' models obtained by the tuning procedure described in the text. The data points correspond to the ATLAS minimum bias measurement at $\sqrt{s}=7$ TeV \cite{ATLASminbias}.} 
\label{fig:tunes}
\end{figure}

As is well known \cite{Perugia,CMS-CR}, the model without CR fails to describe the rising $\langle p_T\rangle(n_{\mathrm{ch}})$. The default and `move' models provide the best description of the data (Figure \ref{fig:tunes}). The `swap' model predicts a smaller slope for $\langle p_T\rangle(n_{\mathrm{ch}})$ than the one observed in data, however the agreement with data is improved in the `swap+flip' scenario.

With the new CR models tuned to data, we reexamine their effect on the top mass. The results are given in table \ref{table:fits2}. We observe that big $m_{\mathrm{top}}$ shifts are disfavored by data. However, even with a reduced CR strength, models that enforce CR in top decays lead to a bigger shift in $m_{\mathrm{top}}$ than the default model, and $m_{\mathrm{top}}^{\mathrm{default}}-m_{\mathrm{top}}^{\mathrm{no-CR}}$ doesn't cover the range of possible $m_{\mathrm{top}}$ values. It is worth noticing that the shifts around $m_{\mathrm{top}}^{\mathrm{no-CR}}$ are not symmetric and we have found that, with the current range of models, it is easier to get a shift towards lower mass than a positive shift. Therefore one could consider assigning an asymmetric uncertainty as far as the CR modeling is concerned.

We conclude this section by noting that if one uses \textsc{Pythia}~8 to derive an uncertainty for the top mass, $m_{\mathrm{top}}^{\mathrm{default}}-m_{\mathrm{top}}^{\mathrm{no-CR}}$ would probably be an underestimate. Taking $m_{\mathrm{top}}$ from the default model as the central value, we observe a maximum shift of around 200 MeV in the positive direction and around $800$ MeV in the negative direction, if all CR models are considered. Restricting oneself to the more sophisticated models that involve a minimization procedure, the maximum deviation from $m_{\mathrm{top}}^{\mathrm{default}}$ is around 500 MeV. We consider the latter to be a realistic estimate for the CR uncertainty on the top mass in view of the existing data and of our current understanding of CR. In the next section we discuss how further constraints on CR models can be established.

\begin{table}[tbp]
\centering
\begin{tabular}{|lccc|}
\hline
Model & $\widehat{m}_{\mathrm{top}}$ [GeV] & $\Delta \widehat{m}_{\mathrm{top}}$ [GeV]  &$\Delta \widehat{m}_{\mathrm{top}}^{\mathrm{rescaled}}$ [GeV]\\
\hline\hline 
CR off						& $168.953\pm 0.024$ & 0                                  & 0\\
default						& $169.135\pm 0.029$ & $+0.182\pm 0.038$ &  $+0.239\pm0.065$  \\
default ERD					& $169.313\pm 0.028$ & $+0.360\pm 0.037$ &  $-0.039\pm0.067$  \\
forced random					& $168.421\pm 0.036$ & $-0.532\pm 0.043$  &   $-0.524\pm0.079$ \\
forced nearest					& $168.712\pm 0.031$ & $-0.241\pm 0.039$  &   $-0.200\pm0.72$ \\
forced farthest					& $168.343\pm 0.036$ & $-0.610\pm 0.043$  &   $-0.515\pm0.077$ \\
forced smallest $\Delta\lambda$  	& $168.823\pm 0.028$ & $-0.130\pm 0.037$  &   $-0.144\pm0.070$ \\
smallest $\Delta\lambda$			& $168.604\pm 0.026$ & $-0.349\pm 0.035$  &   $-0.145\pm0.072$  \\
swap						& $168.843\pm 0.031$ & $-0.110\pm 0.039$  &    $+0.273\pm0.080$ \\
move						& $169.016\pm 0.038$ & $+0.063\pm 0.045$  &   $+0.239\pm0.084$ \\
swap+flip (excl. junc.)			& $169.003\pm 0.038$ & $+0.050\pm 0.045$  &   $+0.196\pm0.084$ \\
swap+flip (incl. junc.)			& $168.902\pm 0.032$ & $-0.051\pm 0.040$  &   $+0.098\pm0.086$ \\
move+flip (excl. junc.)			& $169.090\pm 0.040$ & $+0.137\pm 0.047$  &   $-0.044\pm0.092$ \\
move+flip (incl. junc.)			& $169.139\pm 0.040$ & $+0.186\pm 0.047$  &   $+0.158\pm0.088$ \\
\hline\hline
\end{tabular}
\caption{\label{table:fits2}Values of $\mtop$ predicted by the different CR models after tuning to existing LHC data. The values of $\Delta \widehat{m}_{\mathrm{top}}$ before and after rescaling are calculated in the same way as in Table \ref{table1}.}
\end{table}

\section{Constraints from future measurements}\label{sec:observables}

The new CR models that we introduced are designed to behave in different ways, which eventually manifest in different predictions for observables. Ultimately we expect that one of these predictions will approximate better what happens in Nature with the rest of the models deviating from observations. Insofar as available data cannot distinguish between different models, one usually takes the spread in their predictions as an estimate for the associated modeling uncertainty on a given observable. In the previous section we showed that after tuning to existing LHC data, the new CR models give different predictions for the top mass that vary within a range of 500 to 800 MeV. Here we propose how more stringent constraints can be established from future measurements at the LHC, thereby reducing the uncertainty on the CR modeling. 

The first place where one expects a manifestation of CR effects are the hadron or more specifically charged particle spectra. As discussed in section \ref{sec:models}, different CR models tend to increase or decrease the string length, either explicitly or implicitly. Since the string length is proportional to the hadron multiplicity produced when the string decays, an increase (decrease) in string length would result in an increase (decrease) in the produced hadron multiplicity. Moreover since the $p_T$ of the string would have to be shared among more (less) hadrons, the hadron $p_T$ would tend to become softer (harder) in models that minimize (maximize) $\lambda$. We remind that with the exception of the `swap' and `move' models, where reconnections happen only if $\lambda$ is minimized, the new CR models are a mixture of the default model, where reconnections happen probabilistically, and of the new scenarios that involve the extremization of some measure.\footnote{The only exception is the `forced random' model, where all reconnections are stochastic.}

\begin{figure}[h]
\centering
\centerline{\includegraphics[height=6cm]{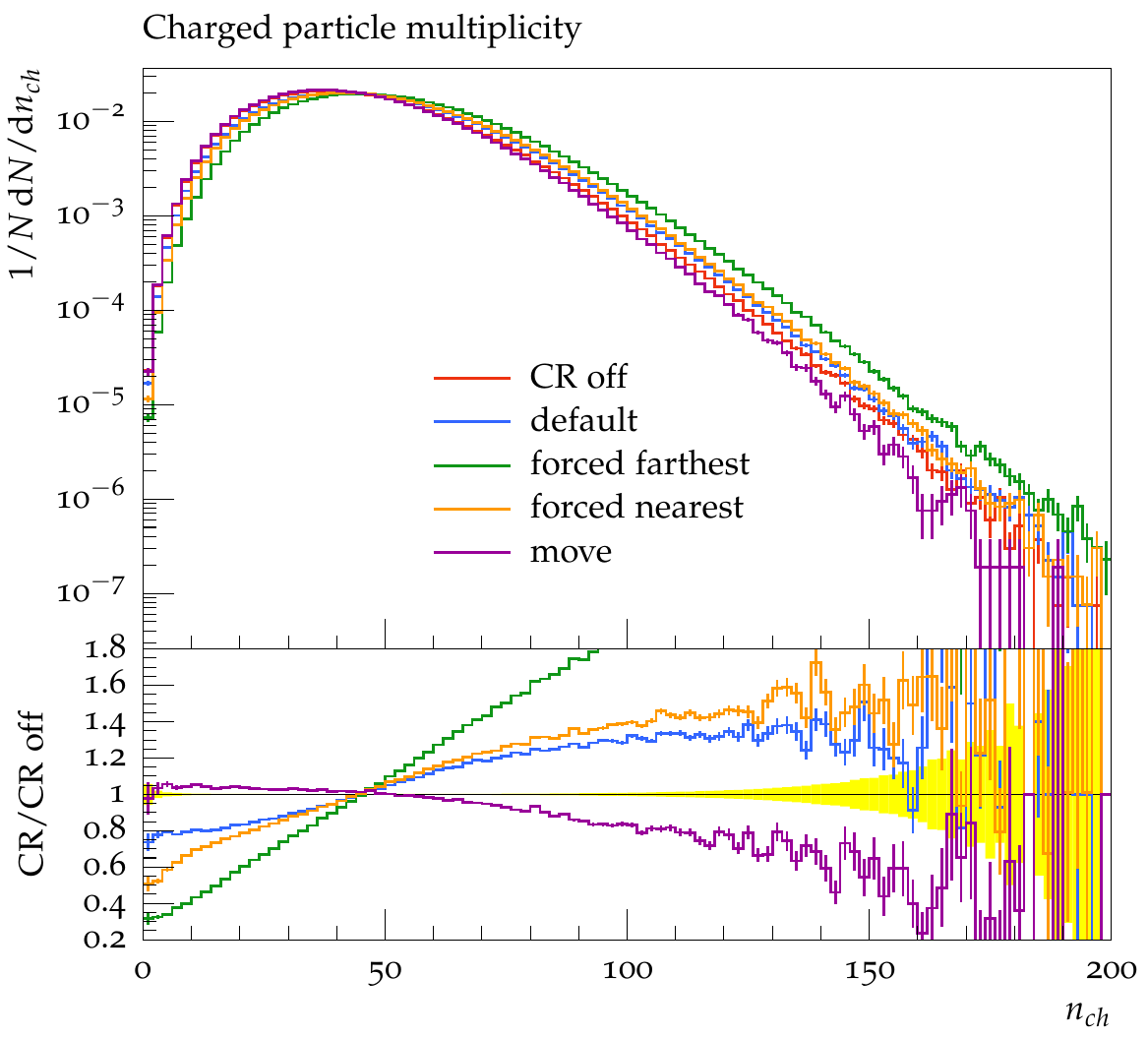}\includegraphics[height=6cm]{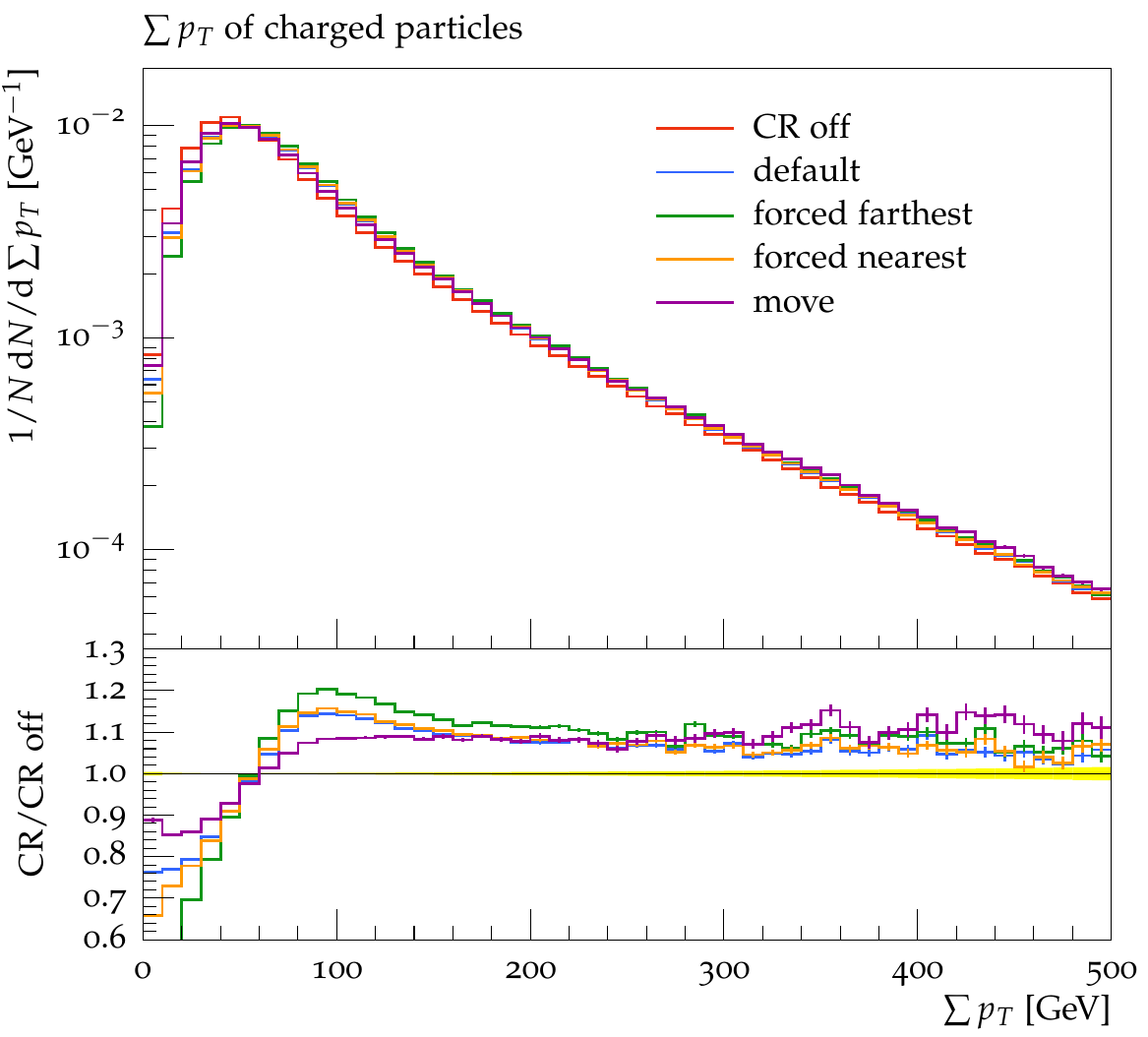}}
\centerline{\includegraphics[height=6cm]{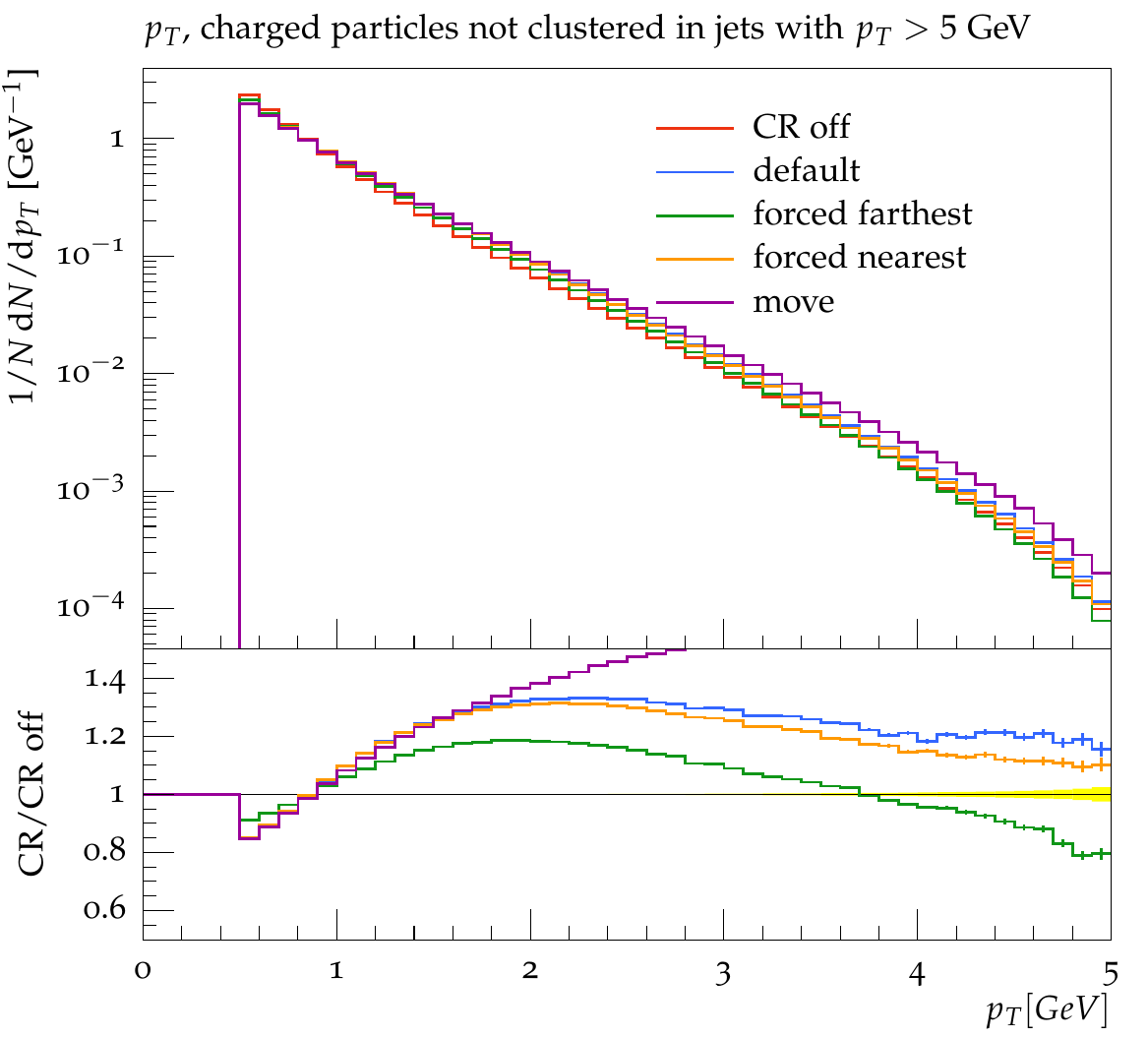}\includegraphics[height=6cm]{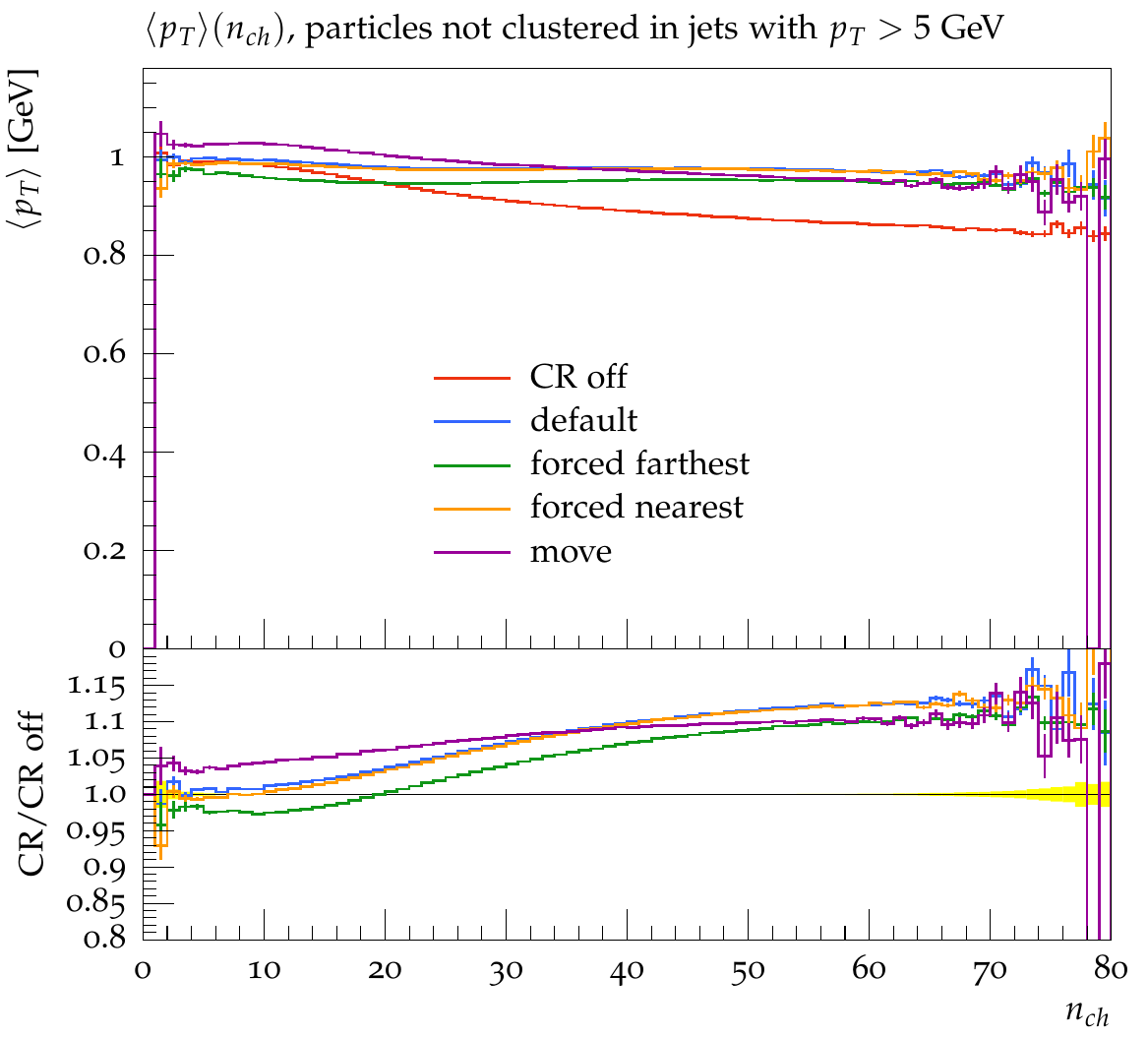}}
\caption{Multiplicity (top left) and scalar sum of $p_T$ (top right) of charged particles not clustered in $b$-jets or the jets that constitute the $W$ candidate. The bottom row shows the $p_T$ (left) and mean $p_T$ per charged particle (right) for particles not clustered in jets with $p_T^{\mathrm{jet}}>5$ GeV.} 
\label{fig:nch}
\end{figure}

In the following we focus on observables reconstructed in semi-leptonic $t\bar{t}$ final states. The multiplicity and scalar sum of $p_T$ of charged particles not clustered in the $b$-jets or in the jets that constitute the hadronically decaying $W$ candidate is shown in figure \ref{fig:nch} for a representative selection of models, using the parameters described in the previous section. The `swap' and `move' models, where only reconnections that minimize $\lambda$ are performed, display the lowest charged particle multiplicity. The multiplicity grows in the models that enforce reconnections involving the top decay products, taking the maximum value in the `forced farthest' model, where reconnections happen between widely separated gluons, thereby increasing the string length. Conversely, the $p_T$ spectrum is softer in the models that lead to an increased string length and harder in models that decrease the string length, in accordance to what we expect. These observables exhibit good discrimination power among the different CR models.

\begin{figure}[h]
\centering
\centerline{\includegraphics[height=6cm]{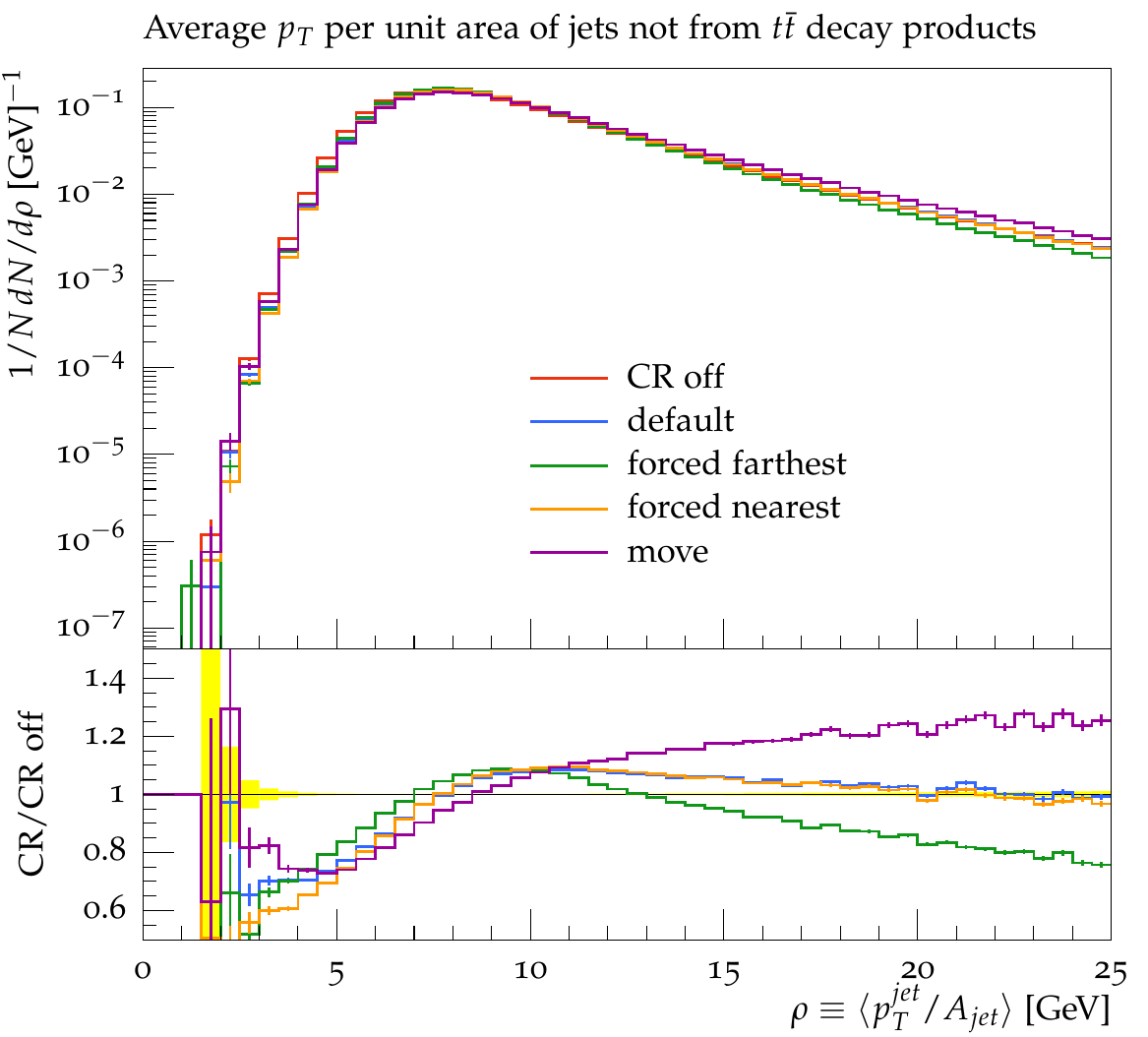}\includegraphics[height=6cm]{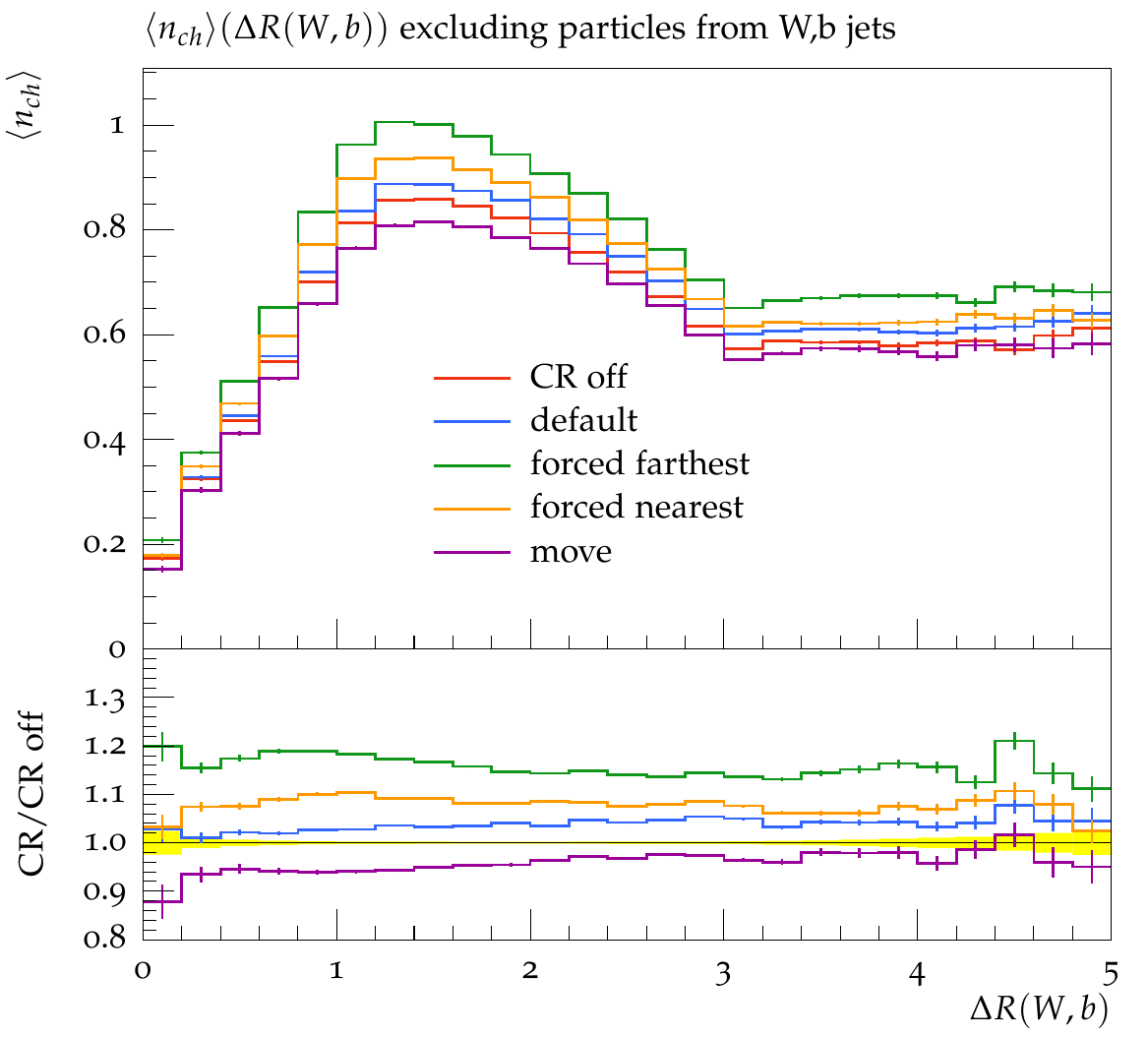}}
\caption{Mean $p_T$ density per unit area (left) calculated using the active jet areas \cite{jetarea} (left) and mean multiplicity of charged particles enclosed in a cone of radius 0.4 around $\vec{p}_{\mathrm{top}}$ as a function of $\Delta R(W,b)$ (right).} 
\label{fig:density}
\end{figure}

The soft component of the underlying event in $t\bar{t}$ events can be accessed by considering charged particles which are not clustered in jets above a certain threshold. Setting this threshold to 5 GeV, we plot the charged particle $p_T$ and mean $p_T$ per particle. These observables also provide a good separation among the different models. Similar information with complementary discrimination power can be obtained by considering the mean $p_T$ density in the event. The latter is given by the mean of the ratio of the $p_T$ of the jets divided by the their corresponding active area \cite{jetarea}. In the calculation one excludes the $b$ jets and the jets that constitute the $W$ boson, to avoid the bias from high-$p_T$ jets. As shown in figure \ref{fig:density}, there is a discrimination among the different CR models at high $\rho$. 

Observables specifically tailored to probe color reconnection effects in $t\bar{t}$ events can also be constructed. A representative example which has proven to have a high discrimination power is the mean number of charged particles clustered in a region between the $W$ boson and the $b$ jet that comprise the top candidate. Due to the $N_c\rightarrow\infty$ approximation of the parton shower, the $W$ boson always decays to a color connected $q\bar{q}$ pair and the $b$ quark from the top decay is always connected to the beam remnant. Therefore the region between the $W$ and the $q\bar{q}$ pair is expected to be sparsely populated with strings. Allowing color reconnections increases the probability that a parton in this region connect to the $b$ or the $q\bar{q}$ string, thus increasing the expected number of charged particles in the intermediate region. The observable that we reconstruct is the mean multiplicity of charged particles that lie within a distance $\Delta R\leq 0.4$ of the axis defined by the 3-momentum of the top candidate on the $\eta-\phi$ plane. We find that different CR models lead to different predictions for $\langle n_{\mathrm{ch}}\rangle$ in the region between the $W$ and the $b$ (figure \ref{fig:density}).

Along a similar path, one can try to detect whether CR induces shifts in the jet axes, due to a rearrangement of the constituent particles. Introducing the multiplicity weighted jet azimuth
\begin{equation}
\overline{\phi}_{\mathrm{jet}}=\phi_{\mathrm{jet}}+\frac{\sum_{i=1}^n\Delta\phi_i}{n},\mbox{with }\Delta\phi_i=
\phi_i-\phi_{\mathrm{jet}} (\mbox{in range } [-R, R])
\end{equation}
where $n$ is the number of particles clustered in the jet, one can calculate the azimuthal difference between $b$ jets that originate from the $t,\bar{t}$ decays: $\Delta\phi(\overline{\phi}_b,\overline{\phi}_{\bar{b}})$. Comparing with the azimuthal difference calculated using the usual jet axes $\Delta\phi(\phi_b,\phi_{\bar{b}})$, could reveal systematic trends in certain CR models to arrange the clustered particles along a certain direction. Therefore one expects that $\langle\Delta\phi(\phi_b,\phi_{\bar{b}})-\Delta\phi(\overline{\phi}_b,\overline{\phi}_{\bar{b}})\rangle$ will change according to the way that reconnections are made in each model. Although a separation between the most extreme models is observed in the case with maximal color reconnection,\footnote{This corresponds to taking the color reconnection range $R_{\mathrm{rec}}=10$ in eq.~(\ref{eq:recProb}) and the color reconnection fraction $\alpha=1$ in the `forced' models.} unfortunately we observe no significant difference among the models after tuning. The same holds for the jet pull observable defined in \cite{jetpull}. 

We conclude this section by drawing attention to the fact that not only are there observables sensitive enough to distinguish between the different CR models, but also these observables are straightforward to measure in the LHC analyses of $t\bar{t}$ events. Constraints from such measurements are essential for reducing the uncertainty related to the modeling of CR.

\section{Conclusions and perspectives}

In this paper we have studied how color reconnection affects the reconstruction of $t\bar{t}$ final states at the LHC using {\sc Pythia}~8. Focusing on the top mass, we argued that the default model may underestimate the CR effects, due to the lack of reconnections involving the top decay products. To remedy this, we introduced two new classes of CR models: toy ones, which are designed to specifically affect the top decay products in a maximal way and more sophisticated ones, which affect all colored partons in the final state. 

We demonstrated that CR models that target the top decay products indeed can produce larger effects than the default model and that CR effects can shift the reconstructed top mass to both higher and lower values with respect to the scenario without CR. This is explained by a combination of two effects: changes in the jet shapes and changes in the relative distance of the jets originated by the top decay products. Both effects result from a modification in the topology of the underlying system of hadronizing strings.

The new models have been tuned using LHC data from $t\bar{t}$ and minimum bias events and the spread in the predictions for the top mass has been used to estimate the uncertainty related to the modeling of CR. Considering the more sophisticated models, we observe a maximum deviation of around $500$ MeV from the prediction of the default model. This provides a realistic estimate of the CR uncertainty related to our current understanding of the CR effects.

We have also proposed how the CR uncertainty can be reduced by future measurements at the LHC. More specifically, we have shown that a series of observables probing the spectra of charged particles in $t\bar{t}$ events is sensitive to CR and can distinguish between different models. For the case of the `swap' and the `move' models, which affect all final states, it would be interesting to explore if complementary constraints could be obtained from underlying event measurements at high-$p_T$ scales using different final states, such as inclusive $b$-jets, $W$+jets, Drell--Yan or multi-jet events \cite{highpt-UE}.

We would like to remark that the observables presented here depend in principle on several of the free parameters that the Monte Carlo uses. Therefore future extensions of this work should involve a more comprehensive tuning effort. We note in particular that the impact parameter dependence, the choice of $\alpha_s$ and the fragmentation parameters directly affect the charged particle multiplicity spectrum and the jet shapes and energy scales respectively, which have been found to play a key role in the study of CR. Therefore studies aiming to constrain the CR models should in principle tune these parameters simultaneously with the CR parameters. It should also be noted that the CR models we have introduced here could be extended with further flexibility if need be.

Also next-to-leading-order matrix elements should be included, improving on the shower approximation. This could lead to some systematic top mass shifts, but presumably with approximately the same mass differences between models. 

The use of purely leptonic variables has been advocated as a possibility to avoid top mass uncertainties related to color reconnection. Recent studies \cite{frixione} underline that this involves theoretical uncertainties of its own, however, and so should be viewed as complementary to the observables we study, rather than as a replacement.

We also note that work is under way to incorporate a more sophisticated CR model in {\sc Pythia}~8, based on the full-color SU(3) multiplet structure, including a more sophisticated treatment of baryon formation and a better handling of the beam remnants \cite{JesperPeter}. This will allow further comparisons and tests.

\acknowledgments
Work supported in part by the MCnetITN FP7 Marie Curie Initial 
Training Network, contract PITN-GA-2012-315877, and in part by
the Swedish Research Council, contract number 621-2013-4287. 
We would like to thank Jesper Christiansen, Judith Katzy and Peter Skands 
for discussions and comments on the manuscript.
S.A. would like to thank Elena Yatsenko for technical assistance with the 
Rivet framework. 

% The bibliography will probably be heavily edited during typesetting.
% We'll parse it and, using the arxiv number or the journal data, will
% query inspire, trying to verify the data (this will probalby spot
% eventual typos) and retrive the document DOI and eventual errata.
% We however suggest to always provide author, title and journal data:
% in short all the informations that clearly identify a document.
%\clearpage

\end{document}